\documentclass{article}

\usepackage{arxiv}
\usepackage[utf8]{inputenc}
\usepackage[T1]{fontenc}
\usepackage{hyperref}
\usepackage{url}
\usepackage{booktabs}
\usepackage{amsfonts}
\usepackage{nicefrac}
\usepackage{microtype}
\usepackage{lipsum}
\usepackage{graphicx}
\usepackage{amsmath}
\usepackage[most]{tcolorbox}
\usepackage{xcolor}
\usepackage{tabularx}
\usepackage{colortbl}
\usepackage[numbers]{natbib}
\setcitestyle{maxauthors=3}

\title{Mapping Scientific Literature with Large Language Models and Topic Modeling}

\author{
Mason Smetana \\
Department of Civil and Environmental Engineering\\
University of Pittsburgh\\
Pittsburgh, PA 15213 \\
\texttt{mrs196@pitt.edu} \\
\And
Lev Khazanovich \\
Department of Civil and Environmental Engineering\\
University of Pittsburgh\\
Pittsburgh, PA 15213 \\
\texttt{lev.k@pitt.edu} \\
}

\begin{document}
\maketitle

\begingroup
\renewcommand\thefootnote{}
\begin{NoHyper}
\footnotetext{
Accepted for publication in \textit{Scientometrics}.  
The final authenticated version is available at:  
\url{https://doi.org/10.1007/s11192-026-05643-9}.
}
\end{NoHyper}
\endgroup

\begin{abstract}
Scientific literature is increasingly fragmented by disciplinary boundaries, specialized terminology, and potentially sparse keyword systems, making it difficult to capture the evolving structure of modern science. This study introduces a large language model (LLM)-driven framework for mapping scientific literature from a topic modeling perspective. The approach is demonstrated on a 20-year corpus of more than 1,500 engineering-related articles published in the \textit{Proceedings of the National Academy of Sciences} (PNAS). A two-stage classification pipeline first assigns a primary thematic category to each article based on its abstract, followed by full-text analysis to identify secondary classifications that reveal latent cross-topic connections within the corpus. Unlike conventional topic models, the LLM-based framework produces semantically interpretable topics while maintaining strong quantitative performance. Comparative evaluation against established topic modeling methods shows higher topic diversity and lower overlap with competitive coherence metrics. Manual validation on a randomly sampled subset of abstracts yields an accuracy of 75.9\%. Additional traditional natural language processing analyses confirm that the generated topics correspond to meaningful linguistic patterns in the corpus. A bipartite network linking primary and secondary classifications further reveals implicit thematic relationships that are not readily observable through abstracts or keyword systems alone. The findings indicate that the framework independently recovers much of the journal's editorial dual-classification structure without prior knowledge of its schema. Overall, the proposed approach offers a powerful tool for mapping science and identifying emerging cross-topic connections in research.
\end{abstract}


\section{Introduction}
Today's research is often dense, highly specialized, and interdisciplinary, making it more difficult for the public to understand. Many people feel poorly informed about significant scientific discoveries, partly because scientists often are not trained to communicate with non-experts \cite{field_public-understanding_2001,luzón_public-communication_2013}. Such a misunderstanding could be partially attributed to the “curse of knowledge” -- a cognitive bias where experts struggle to communicate with non-experts due to their deep understanding of a subject, which can lead to the use of overly complex language and jargon \cite{sharon_measuring-mumbo_2014}. This can confuse those without specialized knowledge or even those conducting cross-disciplinary research \cite{guo_personalized-jargon_2024}. For example, engineers might refer to a “100-year flood” without explaining it as a flood with a 1\% annual chance, which can cause unintended misinterpretation among readers \cite{bruin_assessing-what_2013}.

Beyond effective communication, it is also necessary to understand how science is produced, framed, and shared, especially in high-impact journals like the \textit{Proceedings of the National Academy of Sciences} (PNAS) \cite{schekman_charting-the_2008,verma_impact-not_2015}. Retrospective trend analyses and bibliometric reviews serve as compelling tools for this purpose, allowing for the assessment of research significance, the tracking of shifting priorities, and the improvement of communication strategies \cite{shiffrin_mapping-knowledge_2004,bubela_science-communication_2009}. Mapping such trends further supports meta-research, or research on research, which seeks to improve how we perform, verify, and communicate scientific work \cite{zawacki-richter_mapping_2016, ioannidis_meta-research_2018}. These analyses ultimately promote transparency, reproducibility, and greater public confidence in science \cite{wallach_reproducible-research_2018, lupia_trends-in_2024}.

Historically, bibliometric mapping has focused on co-word, co-authorship, and citation networks to identify thematic trends and relationships within scientific literature, especially with easy-to-use tools like VOSViewer \cite{vaneck_10}. While these networks offer powerful visualization and structuring for literature mapping, especially for indicating \textit{who} is performing research through co-author collaboration, the keyword or abstract-based analyses (\textit{what} is happening in a field) can be limited by sparsity and subjectivity. These methods must be supplemented with human judgment and theoretical grounding to avoid superficial or misleading interpretations \cite{donthu_21}. 

Similarly, statistical topic modeling has been very prevalent in science mapping, which enables the automatic identification of themes within a large document corpus without explicit supervision \cite{benz_mapping-the_2025}. However, traditional methods, such as Latent Dirichlet Allocation (LDA), often struggle with the complexity and diversity of published literature. Prevailing challenges include text sparsity, specialized terminology, and lack of contextual understanding in scientific texts that lead to information loss and reduced interpretability \cite{mimno_optimizing_2011, li_scitopic-enhancing_2025}. Conventional topic models also tend to present topics as collections of words, the “tea leaves,” which complicate analysis and often require adept understanding for validity \cite{hernandez_leveraging_2025,pham_topicgpt_2024}.

While more recent deep embedding methods, such as BERTopic \cite{Gro22}, have been proposed to better capture semantic relationships and address these issues by incorporating contextual information, they still face similar challenges in terms of interpretability and the need for qualitative human validation \cite{hoyle_is-automated_2021}. On the other hand, with the recent advancement of large language models (LLMs), there is growing potential to reframe how scientific knowledge is synthesized and disseminated \cite{markowitz_from-complexity_2024,zhang_scientific-large_2025}. Newer approaches, such as combining document clustering algorithms with transformer-based text representations, have shown promise in effectively grouping texts by their semantic meanings, offering a deeper understanding of thematic structures \cite{weng_22}. 

Despite impressive advancements in modern science mapping with LLMs \cite{pham_topicgpt_2024,theo_2025,mu_large_2024}, many of these approaches still rely on outdated evaluation metrics, topic seeding, and unreliable methods to determine an optimal number of topics. Relative to classical topic models, the correlation of automated evaluations (e.g., coherence, diversity) may not directly apply to newer neural-based models \cite{wu_survey_2024}, yet they are still reported in many recent studies. Furthermore, the selection of an appropriate number of topics for topic modeling often relies on reasonable guesses or perplexity minimization, both of which necessitate significant time and effort for sensitivity analysis \cite{zhao_a-heuristic_2015}. Finally, other than the inherent outlier detection in the BERTopic algorithm \cite{Gro22}, many novel approaches do not propose techniques to filter noisy documents. 

Ultimately, the current methods for mapping scientific literature, from static journal categories to sparse keyword tagging and topic modeling, could significantly benefit from the analytical capabilities of modern language technologies. This study introduces a hybrid framework that combines the contextual understanding of LLMs with traditional machine learning (ML) and natural language processing (NLP) techniques to autonomously map the thematic landscape of scientific reserach. The proposed approach analyzes a 20-year collection of 1,519 articles classified under the PNAS Engineering subcategory and aims to address the following:

\begin{itemize}
    \item Identification of major thematic categories without relying on prior (seed) topics, including estimation of an \textit{optimal} number of topics.
    \item Detection of noisy or less dominant articles that can be assigned to an “Other” category to improve the overall quality of the thematic structure.
    \item Evaluation of classification consistency in the absence of ground truth, followed by full-text analysis to identify secondary themes and cross-topic relationships.
\end{itemize}

To address these objectives, a two-stage classification pipeline was developed. First, abstracts are used to establish each article's primary thematic category. The full text is then analyzed to identify secondary themes, enabling the detection of implicit methodological applications and cross-topic relationships within articles. This study contributes (1) an LLM-based framework for topic discovery in scientific literature, (2) a comparative evaluation with traditional topic modeling approaches, and (3) a full-text analysis method for identifying cross-topic relationships within research articles.

\section{Related Works}

\subsection{Science mapping in PNAS}
Previous reviews of the PNAS journal have provided valuable insights into its evolution and topical structure. Mane and B\"orner \cite{Man04} applied burst detection and co-word association maps to identify emerging trends from 1982 to 2001, highlighting the journal's growth and increasing diversity. Around the same time, Boyack \cite{Boy04} used a range of data mining and analytical techniques to further characterize the journal’s evolving content. Since then, subsequent studies have transitioned toward exploring the journal's disciplinary structure. For example, Airoldi et al. \cite{airoldi_reconceptualizing-the_2010} revisited the prescribed classification system, using statistical ML models to analyze the words in abstracts and the co-occurrence of references. They argued that the existing categories do not fully capture the extent of interdisciplinary research, suggesting that the number of articles classified under multiple categories is only a small fraction of what it should be.

Regarding disciplinary focus, several researchers have noted a prevelance of high-impact journal publications in life science research areas (e.g. molecular biology, genetics, immunology, and nueroscience). Ding et al. \cite{ding_disciplinary-structures_2018} compared the disciplinary structure of PNAS to other high-impact journals, namely \textit{Nature} and \textit{Science}, by analyzing publications from 2004-2006 and 2014-2016. Their analysis revealed that PNAS publications were highly concentrated in biology and medicine, making it less discipline-diverse than the other journals. Xie et al. \cite{xie_feature-analysis_2018} explored collaboration patterns among the biological, physical, and social sciences based on over 50,000 PNAS articles published from 1999 to 2013. They found that many authors, particularly in the physical and social sciences, were engaged in interdisciplinary work and that such research was not limited to highly prolific authors.

More recently, Milojević \cite{milojević_nature-science_2020}, also investigating similar high-impact journals, confirmed this non-uniform disciplinary coverage by automatically classifying articles from 2005 to 2015 into fourteen broad research areas. The study showed that disciplines like bioscience and medicine tend to be well-represented; however, other important fields, such as engineering, mathematics, and computers, each make up less than 1\% of contributions. These findings suggest that the journals' structure may not reflect a truly multidisciplinary scope, leaving necessary fields underrepresented and their trends poorly understood.

\subsection{GenAI for Scientific Knowledge Synthesis}
Recent advancements in generative artificial intelligence (GenAI), especially LLMs, offer capable new tools for accessing and synthesizing scientific knowledge \cite{markowitz_from-complexity_2024,zhang_scientific-large_2025}. These models can process and analyze vast amounts of text, making them valuable for assisting literature reviews, extracting findings, and generating hypotheses across disciplines \cite{thapa_chatgpt-bard_2023}. Derivative transformer models, such as those in the Generative Pretrained Transformer (GPT) family, excel in understanding and generating human language, with wide-ranging applications in summarization, translation, and beyond \cite{petroșanu_tracing-the_2023}. The potential for these tools to revolutionize science communication and accessibility is immense, offering a promising future for the field.

In practice, GenAI's application to scientific literature has primarily concentrated on a specific challenge: improving public accessibility through lay summarization \cite{shyr_leveraging-artificial_2024}. Initiatives like Paper Plain and numerous studies have demonstrated that LLMs can effectively simplify complex texts to enhance user comprehension and engagement \cite{markowitz_from-complexity_2024, guo_automated-lay_2021, august_paper-plain_2023}. This work is valuable and has established a strong precedent for using AI to bridge the gap between scientists and the public. However, this prevailing focus on summarization has left the deeper analytical capabilities of these models for meta-research largely underexplored, presenting an intriguing area for additional investigation.

While making science accessible is one goal, using AI to understand the very structure and evolution of science is another equally critical challenge. The same contextual understanding that allows an LLM to simplify text also allows it to identify thematic connections, map disciplinary overlaps, and detect emergent trends. While LLMs demonstrate increasing proficiency and technical reasoning in domain-specific tasks \cite{Hua23}, their capabilities can be further enhanced by integrating them with established ML and NLP methods to analyze large-scale textual data. A conjunction of such techniques makes it possible to systematically extract, classify, and synthesize complex scientific knowledge \cite{peng_a-study_2023, dagdelen_structured-information_2024}, representing a significant but still underexploited opportunity to advance research understanding.

\subsection{The Evolution of Topic Modeling and Persistent Challenges}
Building upon the concepts of Bag-of-Words (BoW) and Term Frequency-Inverse Document Frequency (TF-IDF), topic modeling has evolved as a powerful tool for uncovering latent thematic structures within large text corpora. Early methods, such as Non-Negative Matrix Factorization (NMF) \cite{lee_learning_1999} and LDA \cite{blei03}, introduced reproducible frameworks for modeling topics as distributions over words, paving the way for the now mature field of latent topic discovery. As NLP techniques have advanced, specifically through the more recent advent of transformer-based models \cite{Vas17}, topic modeling has also evolved to incorporate contextual embeddings (e.g., BERTopic), allowing for more nuanced and semantically rich representations of text \cite{Gro22}. These techniques have been applied to a wide range of domains, from social media analysis \cite{egger_a-topic_2022} to medicine \cite{ma_ai-powered_2025}, and almost every discipline in between.

Despite the widespread adoption of topic modeling algorithms in practice, there remain challenges in interpretation and usability. \cite{benz_mapping-the_2025} recently provided a comprehensive guide for the use of LDA and embedding-based topic models, highlighting the importance of careful preprocessing, parameter tuning, and validation. Their work argues that these techniques are invaluable for science mapping, but also emphasizes the importance of understanding how hyperparameters should be manipulated and evaluated. Many algorithms have a reliance on automated word co-occurrence metrics, such as $C_V$ and $UMass$ coherence \cite{roder_exploring-the_2015}; however, recent research has questioned the reliability of these metrics. Although these measures were designed to correlate with human evaluations of topic coherence, studies have shown that high coherence does not necessarily guarantee that topics are labelable or interpretable \cite{doogan_topic-model_2021}. While they can provide some insight into the internal consistency of topics, they should not be taken literally \cite{hoyle_is-automated_2021}.

These concerns raise questions about whether these models are genuinely performing well on metrics or merely reproducing them without reflecting actual interpretability in real-world applications. Moreover, another challenge with topic modeling is the requirement of determining an optimal number of categories for a dataset -- either via post-analysis evaluation or predisposed intuiton. \cite{zhao_a-heuristic_2015} suggest a heuristic approach to find this optimal number based on perplexity, a metric that assesses how well a model can predict new documents, through an elbow method technique common in unsupervised clustering. This approach has immediate functionality with automatic programming capability, but it has been found that perplexity does not accurately reflect the quality of the topics discovered, often contradicting human judgment \cite{wu_survey_2024}.

Beyond the methodological challenges involved with topic modeling, the evaluation of topic models for data with no ground-truth remains generally inconclusive in research communities \cite{wu_survey_2024}. However, aside from the aforementioned utility of modern GenAI for simplifying texts through lay summarization, LLMs have been significantly praised for their ability to simulate human-like reasoning. This capability has been leveraged in recent studies to enhance topic modeling frameworks, such as using LLMs to generate more coherent and interpretable topics or to assist in labeling topics with meaningful descriptions \cite{hernandez_leveraging_2025,ma_ai-powered_2025}. Classical methods like LDA and NMF often struggle with grasping subtle meanings and require extensive data preparation, making them labor-intensive \cite{kaur_moving-beyond_2024}. In contrast, LLMs like GPT-3 improve adaptability to different contexts, require significantly less preprocessing, and their outputs are often preferred by users over traditional categories \cite{mu_large_2024}.

Amidst the `hype' around chatbots like ChatGPT, \cite{pham_topicgpt_2024} introduced a highly cited LLM-based topic modeling manuscript: TopicGPT. This approach uncovers topics in texts through a prompt-based hierarchical framework, offering enhanced interpretability and interactive adaptability compared to traditional topic models. The algorithm works by first prompting an LLM to generate an initial set of high-level topics from a representative sample of the corpus, then refining this list by merging near-duplicates, and finally assigning these refined topics to the remaining documents. Since then, other studies have proposed similar LLM-based topic modeling approaches, such as those found in \cite{theo_2025} and \cite{mu_large_2024}. 

These methods show promise for improving topic modeling with modern technologies, but they also raise questions about the consistency of LLM-generated outputs, especially given the potential for hallucinations and variability in responses \cite{Far24}. Additionally, these studies offer limited resolution on determining an \textit{optimal} number of topics or finding outliers in a corpus with no ground-truth. For these reasons, the methodology developed in this study seeks to address these limitations by introducing a structured, iterative classification protocol that incorporates multiple runs and agreement scoring to enhance the stability and reliability of LLM-generated thematic categories. 

\subsection{Journal Structure and Engineering in PNAS}
Over the past two decades, PNAS has published nearly 70,000 original research articles, averaging more than 3,000 articles per year \cite{Nat25}. The journal's scope includes the physical, social, and biological sciences, with authors being able to dually select a major and minor subcategory upon submission \cite{airoldi_reconceptualizing-the_2010}. As shown in the \hyperref[tab:pnas_pubcount]{Appendix, Table \ref{tab:pnas_pubcount}}, the biological sciences account for the majority of publications, while the social sciences contribute the fewest. Neuroscience is the most published subcategory, whereas Demography is the least, with publications only appearing since 2022. 

Between 2005 and 2024, over 1,500 articles fell into the Engineering category, a subset of the physical sciences, representing just 2\% of all PNAS publications. In 2024 alone, the journal published over 3,200 original research articles, of which 140 (4.4\%) were classified under Engineering. While modest, this proportion reflects the journal's historical focus on the biological sciences. However, publication rates within the physical sciences, including Engineering, have increased over time, while output in the biological sciences has decreased relatively. This shift suggests a gradual broadening of the journal’s disciplinary reach.

The existing major and minor dual-classification schema further reveals implicit interdisciplinary connections and diversity in PNAS research. Within the engineering collection nearly 44\% of articles were also categorized under a biological science subfield, with Applied Biological Sciences (11.5\%), Medical Sciences (10.8\%), and Biophysics and Computational Biology (6.3\%) being the most common. In contrast, co-classification with social science subfields was rare, occurring in fewer than 1\% of articles.

A preliminary keyword analysis of this engineering subcategory further reinforces its inherent diversity and highlights limitations of traditional analytical approaches. Each author selects at least three unique keywords per submission; a majority of which were used in less than one percent of the articles (15 or fewer occurrences). Dominant terms like “microfluidics,” which appear in 4.8\% of publications (72 occurrences), and “drug delivery” (2.9\%) show consistent representation. In contrast, emerging terms such as “machine learning” (1.5\%) have only gained prominence in the last four years. This extreme sparsity makes it challenging to identify significant research themes using keywords alone. Without additional context or domain expertise, the relevance of many terms (e.g., “fatigue,” “aptamer,” “crispr”) remains ambiguous. These challenges suggest that traditional keyword analysis or topic modeling would require substantial, often impractical, expert oversight.

\section{Methodology}
\begin{figure}[h]
\centering
\includegraphics[width=0.90\textwidth]{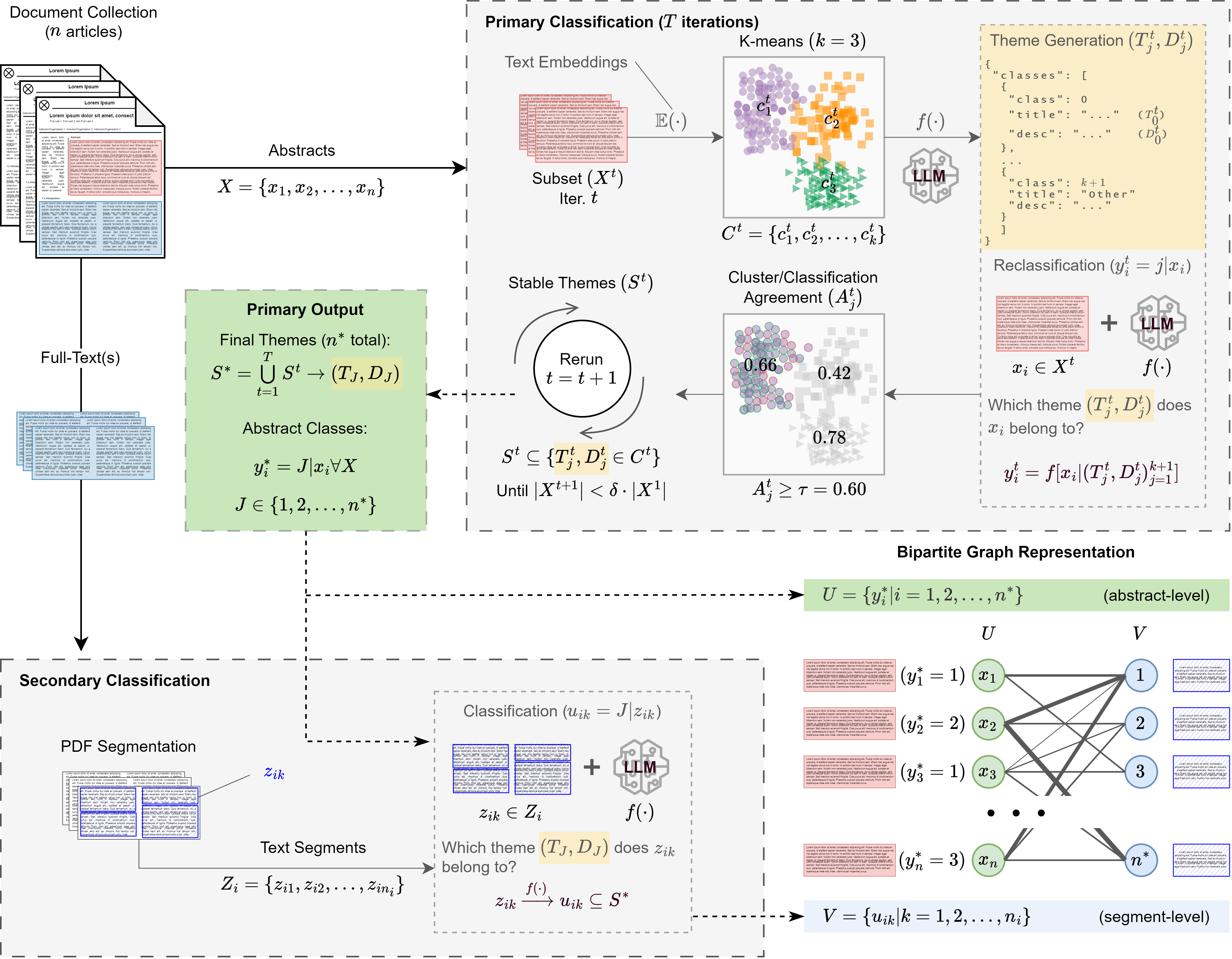}
\caption{Overview of methodology.}
\label{fig:method}
\end{figure}

A general overview of the two-stage LLM classification pipeline is presented in \hyperref[fig:method]{Fig. \ref{fig:method}}. The upper portion of the figure depicts the iterative primary (abstract) classification phase. This process includes clustering texts via high-dimensional embeddings, LLM-generated thematic categories, a structured reclassification protocol, agreement score calculations, and recursive refinement of unstable clusters. The remaining lower portion of the figure illustrates the secondary (full-text) classification phase, which includes the multi-label classification framework and bipartite graph representation to capture connections between abstracts and segmented body texts. Detailed explanations of each step in primary and secondary processes are provided in the following sections.

\subsection{Primary Abstract Classification}
Let $X=\{x_1,x_2,\dots,x_n\}$ be a set of abstracts and $X^t$ be a subset of $X$ in iteration $t$ ($X^t\subseteq X$). Initially, at iteration $t=1$, the subset will equal the complete set of abstracts ($X^1=X$). To construct the data set, the original research articles were programmatically retrieved under the PNAS “Engineering” category by querying the Crossref repository, which yielded 1,493 articles (99.1\% coverage). To broaden the analytical scope, 26 additional commentary and perspective articles were included, resulting in a total of 1,519 documents.

The proposed framework is a hybrid unsupervised approach, combining embedding-based clustering with LLM-guided refinement. This pipeline leverages the strengths of both paradigms:

\begin{itemize}
    \item Clustering: Generates initial K-means clusters from dense embeddings (\textit{text-embedding-3-small}), providing a data-driven foundation.
    \item LLM Refinement: Uses \textit{GPT-4o-mini} to annotate, validate, and refine these clusters, ensuring semantic interpretability.
\end{itemize}

\subsubsection{Clustering}
Each abstract $x_i \in X^t$ is first encoded into a dense vector representation in $\mathbb{R}^d$ using an OpenAI's \textit{text-embedding-3-small} model $\mathbb{E}$ \cite{Nee22,Ope24}. The recursive process begins by clustering the current set of abstracts, $X^t$, into $k$ clusters using the K-means algorithm. A similar procedure is performed in \cite{sme24}, which results in a partition of $X^t$ into a fixed $k$ number of cluster groups $C^t=\{c_1^t,c_2^t,\dots, c_k^t\}$ for iteration $t$: 

\begin{equation}
C^t = \text{KMeans}\Big[\mathbb{E}\left(X^t\right), k \Big]
\end{equation}

This approach is error-agnostic in the sense that, by fixing the number of clusters $k$ in each iteration, the method maintains consistency and avoids the need for external optimization criteria (e.g., inertia, silhouette scores) \cite{Shi21}. However, this also assumes that an appropriate $k$ can be predefined or selected heuristically for the data set at each recursion step. Future implementations may explore adaptive methods to dynamically select $k$, although such approaches would require additional convergence checks to ensure stability across recursive iterations.

\subsubsection{LLM-Generated Thematic Categories}
OpenAI's \textit{GPT-4o-mini} \cite{Ope241} model assists in the generation of thematic categories based on the results of the previous clustering step. For simplicity, the model is denoted as $f(\cdot)$, where input texts are processed to generate outputs through probabilistic predictions across multiple, connected transformer blocks \cite{Vas17}. This generation process can be expressed as:

\begin{equation}
    f(\cdot) = g \Big[ h(\cdot, W_h), W_g \Big]
\end{equation}

Where $h(\cdot, W_h)$ refers to the transformer blocks, constituting the LLM architecture's core, and $W_h$, $W_g$ represent the learned weights of the transformer and output layers, respectively. These layers allow the model to autoregressively generate outputs based on the input \cite{Bro20}. The prompts in the \hyperref[sec:app_A]{Appendix} guide $f(\cdot)$ to annotate a theme label $T_j^t$ and description $D_j^t$ for each cluster $c_j^t \in C^t$, similar to that proposed in \cite{theo_2025}.

\subsubsection{Reclassification}
The LLM is then prompted to reclassify each abstract $x_i \in X^t$ into one of the $k$ categories. $f(\cdot)$ performs this reclassification by evaluating alignment between each abstract $x_i$ and the set of themes ($T_j^t$, $D_j^t$), resulting in a predicted category $y_i^t$:

\begin{equation}
    y_i^t=f \Big[ x_i \mid (T_j^t, D_j^t)_{j=1}^k \Big]
\end{equation}

A structured reclassification protocol was implemented to enhance robustness and reduce susceptibility to hallucinations \cite{Far24}. For each iteration $t$, the model was prompted to reclassify every abstract five times using the previously generated topic labels $T_j^t$ and descriptions $D_j^t$. A classification was considered valid only if at least three of the five predictions agreed on the same topic label. This “3-out-of-5” consensus criterion increases confidence and reproducibility by minimizing the likelihood of predictions driven by chance, given the inherent stochasticity of LLM outputs.

Additionally, to mitigate sequence-related biases due to the autoregressive nature of transformer-based models \cite{Rad19}, abstracts are shuffled in random order across iterations. Since these models rely on prior tokens to generate subsequent output, simultaneous classification of multiple inputs could theoretically result in in-context interference \cite{Fer23}. A safeguard against such sequence-dependent prediction artifacts was introduced by randomizing input batches and repeating sub-iterations. Together, these precautions ensure that the reclassification process is stable, reducing the influence of transient context effects or hallucinated outputs.

\subsubsection{Agreement Score}
To realize stability, a computed score then quantifies the alignment between K-means clustering and LLM classification. For each cluster $c_j^t$, the agreement score $A_j^t$ is defined as the proportion of abstracts $x_i \in c_j^t$ for which the LLM-assigned category $y_i^t$ matches the original cluster label $j$. An indicator variable $\lambda_i$ is introduced, equal to 1 if $y_i^t=j$ and 0 otherwise:

\begin{equation}
A_j^t = \frac{1}{|c_j^t|} \sum_{x_i \in c_j^t} \lambda_i
\end{equation}

A high agreement suggests that the LLM's understanding of the thematic category is consistent with the original embedding space partition. Clusters are considered stable when $A_j^t \geq \tau$, where $\tau$ is a predefined threshold empirically set to 0.60 based on previous benchmark evaluations of embedding-based clustering accuracy \cite{Mue23}. A cluster is accepted into the final stable classification set $S^t$ if the agreement score $A_j^t \geq \tau = 0.60$, meaning that the LLM reclassification aligned with the original K-means cluster in at least 60\% of the abstracts:

\begin{equation}
S^t \subseteq \{ c_j^t \in C^t | A_j^t \geq \tau\}
\end{equation}

\subsubsection{Recursive Refinement and Termination}
Abstracts belonging to unstable clusters (i.e., those below the $\tau$ threshold) are carried forward for re-clustering and reclassification and excluded from the stable set $S^t$. These abstracts then form a new subset $X^{t+1}$, which is recursively reprocessed in the next iteration:

\begin{equation}
X^{t+1} = X^{t} \subseteq \{c_j^t \in C^t | A_j^t < \tau \}
\end{equation}

The recursion continues until the subset of remaining abstracts is less than a termination threshold $\delta=0.10$, defined as a fraction of the size of the initial dataset $X^1$ (Eq. \ref{eq:termination}). Any abstracts that remain unclassified after $T$ iterations may indicate that the model has exhausted primary theme generation. This collection can be grouped into a general “Other” category, acknowledging that they either lack a coherent thematic alignment with the majority of themes or represent outliers.

\begin{equation}
|X^{t+1}| < \delta \cdot |X^1|
\label{eq:termination}
\end{equation}

\subsubsection{Final Output}
A cumulative set of topics is compiled after the termination criteria are met, where each abstract is associated with an LLM-generated label $T_j^t$ and description $D_j^t$. The final output ($S^*$) consists of all stable clusters collected over $T$ iterations (Eq. \ref{eq:final_output}). Each stable cluster $c_j^t$ must then be remapped to a new global index $J$ as a result of the recursive process. This remapping creates consistent, non-overlapping output for $n^*$ final themes $C_J^* \in S^*$ where $J \in \{1,2, \dots, n^*\}$.

\begin{equation}
S^*= \bigcup_{t=1}^{T}S^t = \{ C_1^*, C_2^*, \dots, C_{n^*}^* \}
\label{eq:final_output}
\end{equation}

\subsection{Topic Modeling Evaluation}
Several topic modeling approaches were implemented as baselines for comparison with the proposed LLM classification method. These included LDA \cite{blei03}, NMF \cite{lee_learning_1999}, and BERTopic \cite{Gro22}. Common Python and Java-based libraries (e.g., Gensim, Scikit-learn, and MALLET) were implemented for LDA and NMF, while BERTopic was executed using its native Python package. Each model was applied to the same corpus of abstracts, with standard preprocessing steps including tokenization, stopword removal, and lemmatization \cite{kaur_moving-beyond_2024}. Moreover, it is well known that topic modeling algorithms are sensitive to the choice of hyperparameters, such as the number of topics and seed Dirichlet distributions, $\alpha$ and $\beta$, in LDA \cite{zhao_a-heuristic_2015,wu_survey_2024}. 

Given these inherent model optimization challenges, the same hyperparameters were applied across LDA and NMF approaches to ensure a fair comparison. Specifically, the number of topics was set to the number of stable topics ($n^*$) for each model, excluding the “Other” category as a result of the proposed approach. Since the HDBSCAN clustering model in BERTopic automatically determines outliers, $n^* + 1$ topics were considered in this algorithm. From a preliminary analysis of LDA models, MALLET tends to perform best over the corpus, with the internal optimization strategy for Dirichlet distribution parameters -- which are not always documented adequately. For consistency, $\alpha$ is equal to $5/n^*$ and $\beta$ is equal to 0.01 for all LDA algorithms.

In addition to classical topic modeling, TopicGPT \cite{pham_topicgpt_2024} and the Principal Direction Divisive Partitioning (PDDP) clustering algorithm \cite{theo_2025} were implemented as LLM-based baselines. The same \textit{GPT-4o-mini} model was used for both approaches, with prompts designed to align with the original methodologies proposed in their respective studies. For TopicGPT, the initial topic generation step was performed on a representative sample of abstracts, followed by refinement and assignment to the full corpus. For PDDP, the partitioning process was applied to the entire set of abstracts, with topic labels generated at each split.

To evaluate the performance of each topic modeling approach, automated metrics including coherence scores ($C_V$ and UMass), topic diversity ($T_D$), and Jaccard similarity were calculated \cite{roder_exploring-the_2015, tran_topic-cropping_2013}. However, these metrics should be interpreted with caution, as they may not fully align with human judgement \cite{wu_survey_2024}. Therefore, a manual inspection of the resulting topics was also conducted to assess their relevance to the engineering domain. This qualitative evaluation involved reviewing the top terms and representative abstracts for each topic to determine whether they formed meaningful thematic clusters, as described in the following section.

\subsubsection{Word Frequency Analysis}
Due to the complications that arise from the automated evaluation of topic models with no ground-truth, an independent qualitative inspection of the resulting topics was conducted to assess topic relevance. To achieve this, word frequency analyses through Bag-of-Words (BoW) and Term Frequency-Inverse Document Frequency (TF-IDF) were employed \cite{Qad19,Sah23}. Identical preprocessing techniques (tokenization, stopword removal, and lemmatization), as described previously, were also utilized for this inspection. The BoW and c-TF-IDF analyses were performed on the abstracts associated with each topic cluster, allowing for the identification of dominant terms that characterize each category.

From an NLP standpoint, TF-IDF is a widely used technique for understanding textual importance and thematic structure across documents \cite{Sah23}. However, standard TF-IDF fails to reflect inter-class distinctiveness when applied to grouped or clustered textual data, as in the case of the topic classes generated through LLM-based classification. To address this, a modified class-based TF-IDF (c-TF-IDF) approach was implemented, which reorients TF-IDF to operate at the group level, quantifying how characteristic a term is within a specific topic cluster relative to all others. This method, introduced by \cite{Gro22}, is particularly useful when categories are generated through unsupervised methods, as it amplifies terms that are disproportionately present in one group over others.

In contrast to Grootendorst's original implementation, which combines c-TF-IDF with dimensionality reduction (UMAP) and clustering (HDBSCAN), this approach uses the classification outputs from LLM-driven topic modeling as predefined groups. Each LLM-derived topic was treated as a single class $c$, and the corresponding abstracts were analyzed to extract term patterns specific to that group. Furthermore, the original PNAS keyword structure was mirrored by preserving both unigrams and bigrams during tokenization. Bigrams often capture domain-specific phrasing (e.g., “gene expression” vs. “gene” and “expression”) and were essential in retaining scientific terminology.

\subsubsection{Manual Validation}
Iterating on the lack of ground-truth labels for this dataset, a manual validation was implemented to assess the accuracy of the LLM-based classification. While the model may show high internal consistency through the “3-out-of-5” consensus criterion, it could still be hallucinating or selecting topics based on its own inherent biases. A subset of 20\% of abstracts, excluding those labeled as “Other” ($n^* + 1$), was selected at random for validation. Over this sample, the authors ranked the top three prospective themes ($T_J, D_J \in S^*$). This validation process was designed to be blind to the original LLM-generated labels, ensuring that evaluation was not influenced by model outputs. 

The authors were provided with only the abstract text and the descriptions of the themes, without any indication of which theme was assigned to each abstract by the model. This approach allows for an unbiased assessment of how well the LLM-derived themes align with human judgment. The top three criteria was considered to evaluate the condition where the model may have assigned a single theme to an abstract that could reasonably fit multiple categories. This also presents an inherent limitation of the primary classification phase, which is designed to assign a single theme to each abstract. To counter this limitation, the secondary full-text classification phase described in the next section allows for multi-label assignments, which can capture the cross-topic nature of the scientific literature more effectively.

\subsubsection{Comparison with PNAS Classifications}
To further evaluate cross-topic connections relative to the new LLM-based classification framework, the results were compared against PNAS's original dual-classification system \cite{Nat25}. Specifically, this comparison assessed the degree to which the new topic groupings aligned with articles originally labeled under two distinct scientific categories by the journal.

Of the original 1,519 engineering-related articles retrieved, 26 commentary and perspective pieces were excluded from this analysis. Additionally, 13 research articles were omitted due to parsing errors, resulting in a working set of 1,493 articles. Among these, 661 articles were initially classified into two PNAS categories according to the journal, yielding a baseline dual-classification rate of 44.3\%. This value serves as a reference point for interpreting the cross-topic character of each LLM-derived class.

Three standard information retrieval metrics were used to contextualize the relationship between the new classifications and the original PNAS dual labels: Precision, Recall, and Lift \cite{Hah17, Ibr22}. Together, these metrics provide a quantitative basis for understanding how the LLM-based classification reproduces cross-topic patterns in PNAS Engineering literature. Precision (Eq. \ref{eq:precision}) measures the proportion of articles within a given LLM-derived class originally dually classified by PNAS. Recall (Eq. \ref{eq:recall}) then quantifies the share of all dually classified articles, from the entire abstract collection, captured by a specific LLM topic. Higher Recall suggests broader coverage of cross-topic material by that topic class. Finally, Lift (Eq. \ref{eq:lift}) provides a normalized measure of how much more (or less) concentrated a topic is in dually classified articles compared to the baseline of 44.3\%. It reflects the relative enrichment of interdisciplinary work in a class.

\begin{equation}
\text{Precision} = \frac{\text{Number of Dually Classified Articles in Class}}{\text{Total Count of Articles in Class}}
\label{eq:precision}
\end{equation}

\begin{equation}
\text{Recall} = \frac{\text{Number of Dually Classified Articles in Class}}{\text{Total Number of Dually Classified Articles in Database } (661)}
\label{eq:recall}
\end{equation}

\begin{equation}
\text{Lift} = \frac{\text{Precision of Class}}{\text{Baseline Rate (0.443)}}
\label{eq:lift}
\end{equation}

\subsection{Secondary Full-Text Classification}
Each article PDF is first preprocessed and converted into plain text using a Python script to enable full-text classification. The extracted text is then parsed into smaller segments (e.g., paragraphs, sentences, or thematic blocks), allowing for more granular analysis. The segmentation strategy varies depending on the document length and structure, but ensures that the full text is divided into manageable units for individual analysis. 

After parsing, each abstract ($x_i \in X$), previously classified into a final thematic cluster $C_J^*$ during the primary classification stage (where $y_i^*=J \mid x_i$), is now linked to a set of full-text segments $Z_i=\{z_{i1}, z_{i2}, \dots, z_{in}\}$. Each segment $z_{ik}$ is then classified across the stable themes ($S^*$) derived from the primary classification. The LLM $f( \cdot )$ maps each segment to a theme in $S^*$ ($T_J,D_J$), producing a segment-level label $u_{ik}$ that may either match the abstract's primary classification $y_i^*$ or correspond to different topic:

\begin{equation}
z_{ik} \xrightarrow{f(\cdot)} u_{ik} \subseteq S^*
\end{equation}

In this secondary classification stage, the LLM was allowed to assign multiple topics to each segment. Unlike the primary classification, which assigned a single topic to each abstract, segments $z_{ik}$ were evaluated against all stable topics $S^*$ identified across $T$ iterations, not only the topic associated with their parent abstract ($y_i^*=J \mid x_i$). Consequently, a segment could be associated with multiple topics. Segments were incrementally evaluated against each stable topic, with an “Other” category as fallback. This yields three possible outcomes for any given segment:

\begin{enumerate}
\item \textbf{Single, Specific Topic Assignment:} The segment is confidently assigned to one topic, with all other classification attempts reverting to “Other.” This indicates a clear, singular thematic focus.
\item \textbf{Indefinite “Other” Assignment:} All attempts return “Other,” suggesting the segment does not strongly align with any predefined topic.
\item \textbf{Multiple Topic Assignments:} The segment is assigned to more than one topic, revealing its cross-topic character or thematic ambiguity.
\end{enumerate}

\subsubsection{Bipartite Graph Representation}
A bipartite graph is a type of graph $G=(U, V, E)$ where $U$ and $V$ are two disjoint sets of nodes and $E$ is a set of edges connecting nodes from one set of nodes to the other, with no edges within the same set \cite{Pez18, Kot15}. In this framework, the bipartite graph represents relationships between the abstract-level and the segment-level classifications. The $U$ set (left-hand nodes) represents the primary classifications $y_i^*$, where each abstract $x_i$ is assigned to one of the final thematic clusters ($C_J^*$). The $V$ set (right-hand nodes) represents the secondary classifications $u_{ik}$ derived from full-text segments $z_ik$ associated with each abstract $x_i$. The graph is defined as:

\begin{align}
U = & \{y_i^* \mid i=1,2, \dots, n^* \} & \text{ (abstract-level classification)} \\
V = & \bigcup_{i=1}^{n^*} \{u_{ik} \mid k = 1,2, \dots, n_i \} & \text{ (segment-level classification)} \\
E = & \{ (y_i^*, u_{ik}) \mid z_{ik} \in Z_i, f(z_{ik}) = u_{ik} \subseteq S^* \} & \text{ (edges)}
\end{align}

Each abstract node $y_i^*$ may connect to multiple segment segment classifications $u_{ik}$.  When $u_{ik}=y_i^*$, the segment aligns with the abstract's primary topic. Otherwise, when $u_{ik} \neq y_i^*$, the segment is associated with a different theme, indicating potential cross-topic relationships. This visualization can be used to explore how knowledge from one subdomain (e.g., biomedical) may connect with other areas of engineering (e.g., material science), potentially highlighting cross-topic trends that could inform new avenues of research.

\subsubsection{Adjacency Matrix}
If multiple topics, for example $C_1^*$ and $C_2^*$, are linked to the same abstract $x_i$ through segment classifications $u_{ik} \subseteq S^*$, these themes are considered connected in the graph. Such edges represent co-classification within the same document and indicate potential cross-topic links. The co-occurrence between themes can be summarized using an adjacency matrix $M \in \mathbb{R}^{n^* \times n^*}$, where each element $M_{pq}$ records how two topics $C_p^*$ and $C_q^*$ appear together in the same article based on segment classifications. Formally, where $1(\cdot)$ is an indicator function that equals 1 if both themes occur within the same article:

\begin{equation}
M_{pq}=\sum_{i=1}^{|X|} 1(\exists k,l \text{ such that } u_{ik}=C_p^*, u_{il}=C_q^*)
\end{equation}

The resulting matrix $M$ summarizes the cross-topic relationships induced by the two-stage classification process: abstract-level nodes ($U$, row in $M$) connect to segment-level nodes ($V$, column in $M$) through edges ($E$), and these edges are aggregated to reveal how frequently topics co-occur within the same article. Since each segment was classified independently from its parent abstract, they could have been classified into a different topic under the same classification schema ($T_J \in S^*$). For example, \textit{GPT-4o-mini} may have classified an abstract ($x_i$) as topic 1 ($y_i=T_1^1$); however, the individual segments ($z_{ik} \in Z_i$) could have been classified into any of the other topics for $u_{ik} \subseteq S^*$ where $u_{ik} \neq T_1$.

\section{Thematic Mapping of Engineering Abstracts}

\subsection{Primary Abstract Classification}
The primary phase of the proposed two-stage approach involved the classification of 1,519 PNAS engineering abstracts ($x_i \in X$), resulting in a stable set of 16 research areas ($S^*$), along with temporal trends and topical structures that were not readily apparent from the journal's existing classification scheme. The iterative procedure initiated by clustering abstracts into $k=7$ using the \textit{text-embedding-3-small} embedding model. \textit{GPT-4o-mini} then generated titles ($T_j^t$) and descriptions ($D_j^t$) for each group, which were used in subsequent reclassification. Abstracts belonging to clusters that failed to meet an agreement score of 60\% were reprocessed in subsequent iterations. The entire procedure converged after $T=9$ iterations, when fewer than 10\% of abstracts remained in the fallback $k+1$ “Other” category. Consequently, 16 of 63 (25\%) total cluster-topic pairs achieved sufficient agreement scores (\hyperref[fig:interim_topics]{Appendix, Fig. \ref{fig:interim_topics}}).

To ensure LLM reliability, each abstract was intermittently reclassified five times against the generated topic set. A “3-out-of-5” confidence criterion was met by 96.2\% of abstracts, with over half receiving consistent labels across all five runs, indicating strong internal consistency in the topic assignment (\hyperref[fig:3of5_criteria]{Appendix, Fig. \ref{fig:3of5_criteria}}). The final, stable set of 16 primary topics captured 90.1\% of all abstracts -- a result driven by the convergence criterion $\delta=0.10$, which effectively reduced the overuse of the “Other” category and promoted more meaningful topic consolidation.

\begin{figure}[h]
\centering
\includegraphics[width=0.5\textwidth]{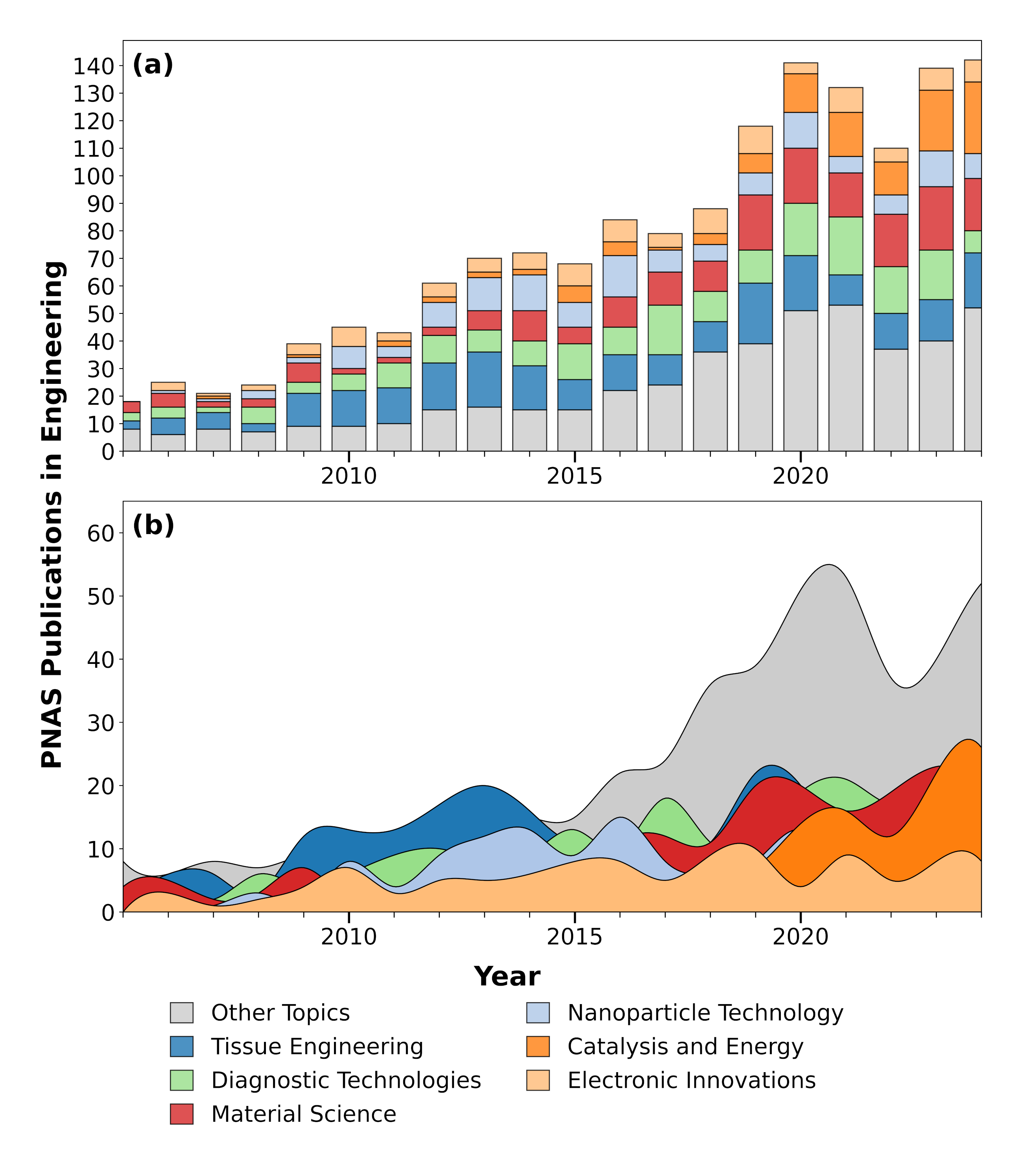}
\caption{Cumulative (a) and overlapping (b) distributions of top LLM-derived topics and classifications for engineering abstracts. Best viewed in color.}
\label{fig:dist1}
\end{figure}

\hyperref[fig:dist1]{Fig. \ref{fig:dist1}} illustrates the (a) cumulative and (b) overlapping temporal trends of the six most prevalent primary topics, which account for more than two-thirds of the classified abstracts. While overall article output has declined since 2012, engineering publications have increased steadily, aligning with the previously discussed trends. The lower plot (\hyperref[fig:dist1]{Fig. \ref{fig:dist1}b}) uses an overlapping area chart to emphasize the changing proportions of different engineering topics over time, making fluctuations and trends within the individual topics more apparent. Together, these subplots illustrate the evolution of engineering-related publications from 2005 to 2024 and their position within the journal's broader topical landscape.

The generated titles $T_J$ (\textbf{bold}), descriptions $D_J$ (\textit{italics}), and key publication trends for the six dominant topics are as follows. The remaining ten topics, also assigned clear titles and thematic descriptions using the same generative procedure, are omitted here due to spatial constraints, as each contributes less than 5\% to the total abstract distribution (See \hyperref[tab:cluster_agreement]{Appendix, Table \ref{tab:cluster_agreement}}).

\begin{itemize}
\item \textbf{Tissue Engineering (and stem cell behavior) -- 16.9\%:} \textit{Study of tissue engineering and mechanotransduction, emphasizing the role of extracellular matrix and stem cell behavior in biomaterial}; Gained interest in 2006 and significantly dominated the journal from 2008 to 2014, with an additional spike around 2019.
\item \textbf{Diagnostic Technologies -- 13.7\%:} \textit{Innovations in diagnostic technologies aimed at enhancing healthcare outcomes through real-time analysis and personalized medicine}; Relatively stable over two decades, with an overall increase in publications within this period. 
\item \textbf{Material Science -- 13.4\%:} \textit{Investigation of mechanical properties at micro and nanoscale levels, focusing on micromechanics and additive manufacturing}; Contributed to many earlier publications from 2005 to 2007. Substantial return after 2015.
\item \textbf{Nanoparticle Technology (for targeted cancer therapy) -- 9.7\%:} \textit{Advancements in cancer treatment utilizing nanoparticle technology for targeted drug delivery and personalized medicine}; Surge of articles surfaced from 2010 to 2017.
\item \textbf{Catalysis and Energy -- 8.1\%:} \textit{Research on catalysis and energy storage technologies, focusing on CO$_2$ conversion and sustainable lithium battery solutions}; After 2019, the subject gained popularity, being one of the most dominant concentrations in recent years.
\item \textbf{Electronic Innovations -- 7.2\%:} \textit{Exploration of cutting-edge electronic and photonic technologies, focusing on 2D materials and flexible electronics}; Consistent contribution to the journal from 2005 to 2024.
\end{itemize}

The primary analysis ultimately identified 16 prominent research topics that were not explicitly defined by the standard PNAS classification system \cite{Nat25}. These generated topics enhance lay interpretability by distilling complex scientific areas into concise language. For example, instead of using fragmented technical phrases such as “stem cell proliferation,” “matrix stiffness,” and “mechanotransduction pathways” scattered across abstracts, the LLM synthesized these elements into a single, coherent topic titled “Tissue Engineering (and stem cell behavior).” Similar titles and descriptions for the remaining fifteen topics can be found in the \hyperref[tab:cluster_agreement]{Appendix, Table \ref{tab:cluster_agreement}}.

Furthermore, the remaining ten topics grouped as “Other Topics” (shaded grey in \hyperref[fig:dist1]{Fig. \ref{fig:dist1}}), gained prominence since 2015. From 2015 to 2024, 369 engineering articles were classified within this emerging group. Among these, primary sub-themes included robotics (soft and bio-inspired) and shape-morphing materials (104 abstracts), biomedical and bioengineering innovations (87 abstracts), fluid dynamics and microfluidics (64 abstracts), and imaging techniques (7 abstracts). The residual 107 abstracts were too thematically diffuse to form cohesive subtopics and did not meet the threshold for standalone primary abstract classification. Overly specific subcategories with little representation (e.g., imaging techniques) would emerge if additional iterations beyond the existing nine were performed.

\subsection{Comparing Topic Models}
The primary abstract classification results were compared with several topic modeling approaches, including implementations of LDA and NMF, TopicGPT \cite{pham_topicgpt_2024}, and the PDDP clustering method \cite{theo_2025}. As shown in \hyperref[tab:topicmodels]{Table \ref{tab:topicmodels}}, the proposed method achieves competitive coherence scores ($C_V$ and UMass) while outperforming several models in topic diversity ($T_D$) and Jaccard similarity.

\begin{table}[h]
\centering
\caption{Comparison of proposed LLM-based classification with other topic modeling approaches.}
\begin{tabularx}{\textwidth}{@{\extracolsep{\fill}} lccrrrr}
\toprule
Method & $n^*$ & Outliers & $C_V \uparrow$ & UMass $\uparrow$ & $T_D$ $\uparrow$ & Jaccard $\downarrow$ \\
\midrule
Proposed & 17 & Yes & 0.446 & -6.416 & \textbf{0.909} & \textbf{6.15E-03} \\
BERTopic & 17 & Yes & 0.491 & -3.933 & 0.765 & 2.90E-02 \\
TopicGPT & 19 & No & 0.406 & -8.290 & 0.900 & 6.54E-03 \\
PDDP & 16 & No & 0.520 & -3.510 & 0.850 & 1.27E-02 \\
Gensim LDA & 16 & No & 0.387 & -3.308 & 0.872 & 1.09E-02 \\
MALLET LDA & 16 & No & \textbf{0.540} & \textbf{-2.328} & 0.747 & 2.49E-02 \\
Sklearn LDA & 16 & No & 0.373 & -2.416 & 0.519 & 9.96E-02 \\
Gensim NMF & 16 & No & 0.348 & -2.251 & 0.503 & 7.78E-02 \\
SKlearn NMF & 16 & No & 0.469 & -3.116 & 0.706 & 3.02E-02 \\
\bottomrule
\end{tabularx}
\label{tab:topicmodels}
\end{table}

The proposed method yields the highest topic diversity ($T_D=0.909$) and the lowest Jaccard similarity ($6.15\times10^{-3}$), indicating that the resulting topics are more distinct and less overlapping compared to those produced by other approaches. The inclusion of an “Other” category in the classification framework may contribute to this separation, whereas most topic models force all documents into predefined topics, which can increase overlap. Even BERTopic, which includes an outlier category, exhibits higher Jaccard similarity, suggesting greater topic overlap. Conversely, the MALLET implementation of LDA achieves the strongest coherence scores, indicating that its topics are more internally consistent according to automated coherence measures. However, its lower diversity and higher similarity scores suggest greater overlap between topics relative to the proposed method.

Among other LLM-based approaches, TopicGPT and PDDP exhibit comparable performance in diversity and similarity metrics. PDDP achieves the strongest coherence scores but lower diversity, while the proposed method marginally outperforms TopicGPT across both coherence and diversity measures. TopicGPT was not provided with seed topics or a predefined number of topics, resulting in 19 discovered topics compared to the 16 topics identified in this study, none of which were labeled as outliers. However, these metrics should be interpreted with caution, as they may not fully align with human judgement \cite{wu_survey_2024}.

\subsection{Topic Validation and NLP Interpretation}
Due to limitations of automated topic modeling metrics in the absence of ground-truth labels, a manual evaluation was conducted to assess the reliability of the LLM-based classification. A random sample of 303 abstracts (20\% of the corpus) was selected for evaluation. For each abstract, the authors independently ranked the three most appropriate themes without knowledge of the model's prediction. \hyperref[tab:manual]{Table \ref{tab:manual}} summarizes the validation results, including the number of abstracts sampled from each topic and the corresponding Top-1 and Top-3 accuracies. Top-1 accuracy indicates cases where the LLM-selected topic matched the highest-ranked human label, while Top-3 accuracy indicates whether the model's prediction appeared among the three human-ranked themes.

\begin{table}[h]
\centering
\caption{Manual validation over random sample (20\% of all abstracts).}
\begin{tabularx}{\textwidth}{@{\extracolsep{\fill}} lrrrr}
\toprule
Topic ($T_{J}$) & No. Abs & Sample & Top-1 Acc. (\%) & Top-3 Acc. (\%)\\
\midrule
01 -- Tissue Engineering ... & 256 & 59 & 37.3 & 50.8 \\
02 -- Nanoparticle Technology ... & 147 & 29 & 24.1 & 72.4 \\
03 -- Catalysis and Energy & 123 & 23 & 73.9 & 100.0 \\
04 -- Electronic Innovations & 110 & 21 & 47.6 & 85.7 \\
05 -- Diagnostic Technologies & 208 & 44 & 43.2 & 72.7 \\
06 -- Material Science & 203 & 42 & 76.2 & 92.9 \\
07 -- Biomedical Engineering & 58 & 17 & 47.1 & 88.2 \\
08 -- Synthetic Biology & 28 & 9 & 88.9 & 88.9 \\
09 -- Bioinspired Robotics & 46 & 13 & 69.2 & 69.2 \\
10 -- Soft Robotics & 43 & 11 & 63.6 & 63.6 \\
11 -- Shape-Morphing Materials & 27 & 6 & 66.7 & 100.0 \\
12 -- Fluid Dynamics & 37 & 8 & 87.5 & 100.0 \\
13 -- Microfluidics Innovations & 25 & 8 & 75.0 & 100.0 \\
14 -- Water Purification & 25 & 4 & 25.0 & 50.0 \\
15 -- Bioengineering Innovations & 23 & 7 & 14.3 & 28.6 \\
16 -- Imaging Techniques & 10 & 2 & 50.0 & 100.0 \\
\midrule
\textbf{Total} & \textbf{1,519} & \textbf{303} & \textbf{52.5} & \textbf{75.9} \\
\bottomrule
\end{tabularx}
\label{tab:manual}
\end{table}

Since many abstracts span multiple engineering subdomains, strict single-label evaluation may underestimate classifier reliability. During manual validation, annotators were allowed to identify multiple themes that could reasonably describe an abstract. Under a single-label evaluation, the model achieved 52.5\% accuracy. When allowing multiple conceptually valid topic assignments per abstract, accuracy increased to 75.9\%. This result reflects the cross-topic nature of the corpus and suggests that some apparent misclassifications arise from taxonomy constraints rather than model error, an issue further explored in the subsequent full-text analysis.

Further investigation of the primary topics was explored through BoW and c-TF-IDF analyses, helping to confirm the shared vocabulary within thematic groups. This interpretation was conducted on the abstract collection containing over 63,000 words, including 48,500 unique unigrams (single words) and bigrams (word pairs). BoW was applied with lemmatization, the process of reducing a word to its base or root form, consolidating the vocabulary to a more manageable size of 6,451 unique unigrams.

\begin{figure}
\centering
\includegraphics[width=0.5\textwidth]{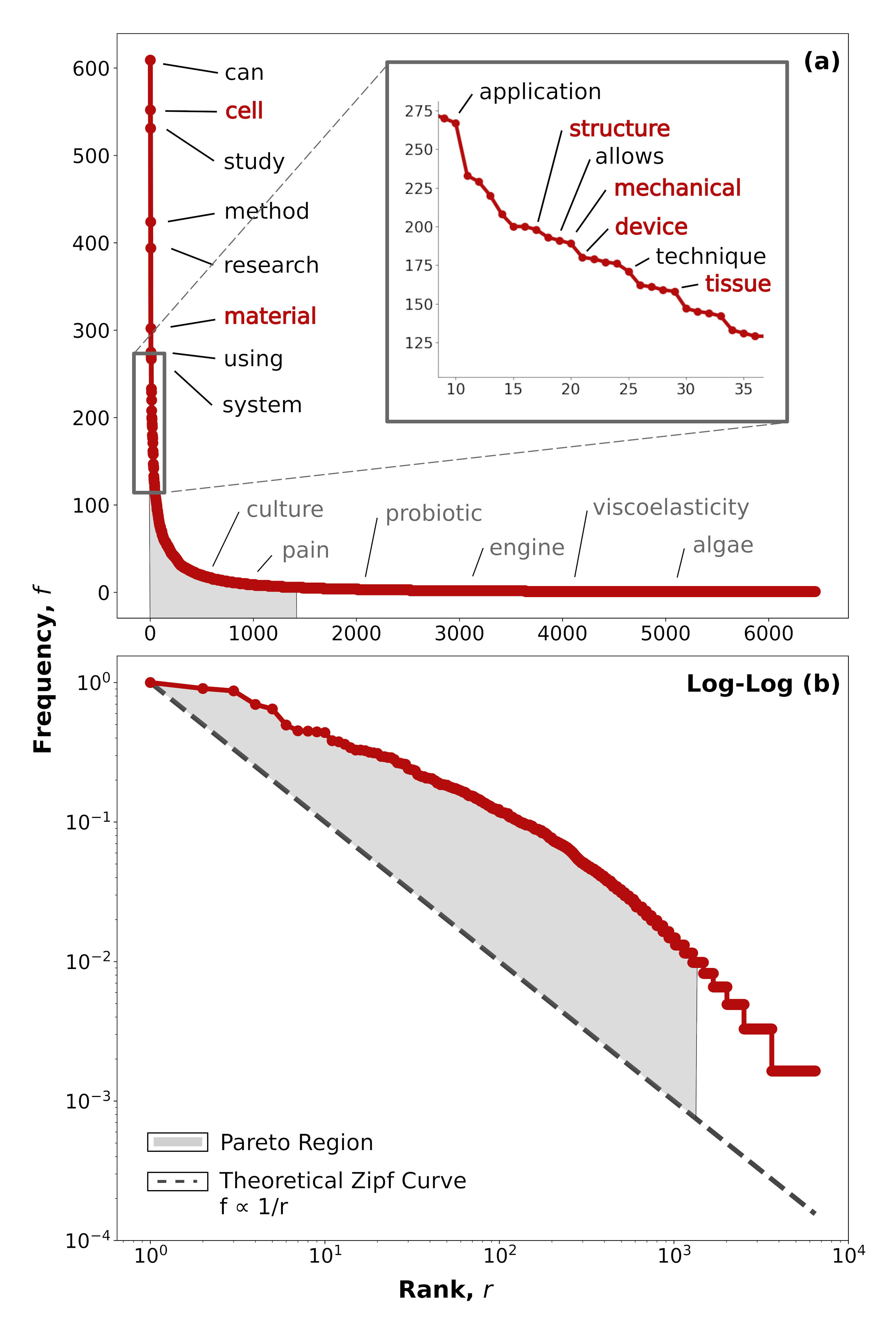}
\caption{Word frequency ($f$) versus rank ($r$) for all abstracts on linear scale (a) and log-log scale with normalized frequency (b) conforming to Zipf's law. The highlighted area under each curve represents the Pareto region, where 80\% of the most frequent words occur in the first ~20\% of rankings.}
\label{fig:zipf}
\end{figure}

A similar notable sparsity, as observed through author-selected keywords, is illustrated through this process. \hyperref[fig:zipf]{Fig. \ref{fig:zipf}} plots word frequency $f$ against rank $r$, demonstrating a clear inverse relationship ($f\propto \nicefrac{1}{r}$) consistent with Zipf's law -- a characteristic feature of linguistic data where the most common word occurs about twice as often as the second, three times as often as the third, and so on \cite{Zip36,Pia14}. This statistical representation also conforms to a head-body-tail structure typical of language following Zipf's law. The head and body regions follow a Pareto-like distribution, where a small fraction of words dominates overall frequency \cite{New05}.

\begin{figure}[h]
\centering
\includegraphics[width=0.99\textwidth]{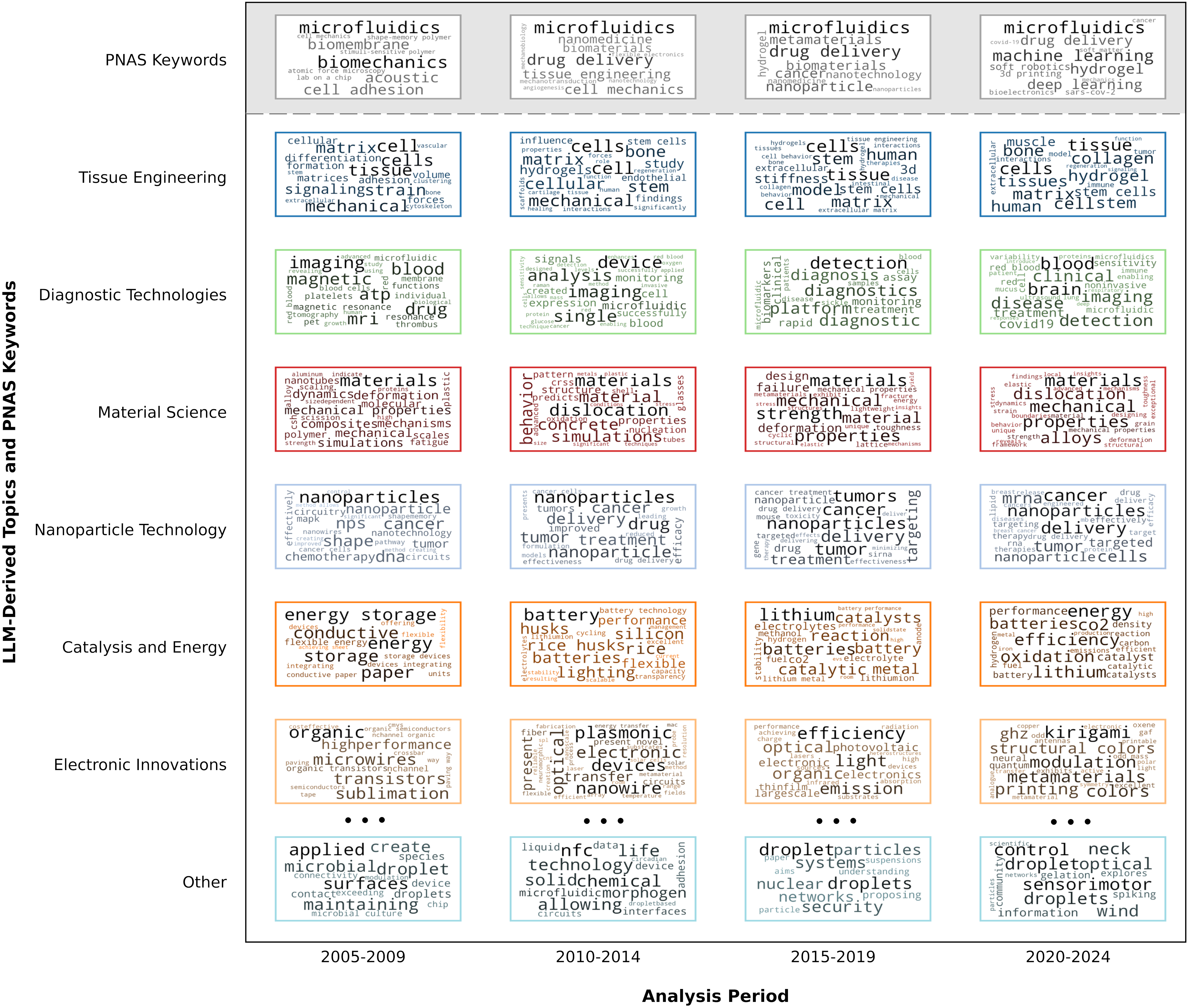}
\caption{Progressive c-TF-IDF word clouds of LLM-derived topic abstract classifications. The x-axis shows four bucketed time periods spanning five years each. The y-axis shows PNAS author-selected keywords over the entire engineering subcategory, the top six research topics from abstract classification, and the final “Other” category. The dots are a placeholder for ten additional topics between “Electronic Innovations” and “Other,” excluded due to spatial limitations.}
\label{fig:ctfidf}
\end{figure}

In this corpus, the top 1,371 ranked words, just 21.3\% of the lemmatized vocabulary, accounted for nearly 80\% of all word occurrences (\hyperref[fig:zipf]{Fig. \ref{fig:zipf}a}, shaded region). Words in this region can be broadly categorized as follows: 50\% are ancillary or general terms (e.g., “study,” “method,” “can”), 8\% are participles (e.g., “increasing,” “enabling”), and the remaining 42\% reflect more domain-relevant vocabulary, including “cell,” “tissue,” “structure,” and “material” (shaded in bold red). Additionally, a characteristic long tail region is prevalent in this distribution: most higher-ranking terms have exceptionally low frequencies. \hyperref[fig:zipf]{Fig. \ref{fig:zipf}b} compares the collection's distribution to an idealized Zipf curve where $f$ is inversely proportional to $r$ (dashed line). The stair-step structure in this region represents when multiple words share identical frequencies (typically $f\leq 5$).

c-TF-IDF was then applied to illustrate the dominant words in the LLM-based primary classification. \hyperref[fig:ctfidf]{Fig. \ref{fig:ctfidf}} presents the top keywords for six of the most prominent LLM-generated topics, aggregated over five-year intervals to highlight temporal dynamics. For comparative context, the figure also includes the top author-selected keywords from the PNAS website, denoted as “PNAS Keywords,” as well as the “Other” category from the abstract classification. The color and size of each word indicate its respective c-TF-IDF score and, therefore, importance, where the darker and larger a word is, the higher the score.

Overall, the resulting term distributions from c-TF-IDF show clear topical separation. For instance, terms like “cell,” “matrix,” and “hydrogel” dominated Tissue Engineering (and stem cell behavior), while Material Science was defined by “dislocation,” “alloys,” and “strength.” Shared terms such as “mechanical” and “stiffness” suggest some overlap, though this technique generally maintains distinct word profiles per topic. In addition to topical separation, topic evolution across time was observed. Within Nanoparticle Technology (for targeted cancer therapy), early years emphasized “dna” and “shape,” while later periods shifted toward “drug delivery,” “tumor,” and “cancer.” Similar transitions appear in other topics, highlighting changes in research focus over the past two decades. 

When distributed into the sixteen LLM-derived topics, the BoW model revealed that higher-ranking terms aligned with dominant c-TF-IDF features. Words like “material” ($f=302$, $r=6$) appeared across all sixteen groups; however, they were primarily concentrated in Material Science ($f_6=155$), suggesting broad relevance with strong topical anchoring (i.e., being much more dominant in one topic compared to others despite its breadth). Similarly, “tissue” ($f=148$, $r=29$) was distributed across only six groups and peaked in Tissue Engineering ($f_1=98$). Approximately 6\% of words were exclusive to a single topic, such as “dislocation” ($f=32$, $f_6=32$, $r=294$), appearing only in Material Science, mirroring later c-TF-IDF patterns (\hyperref[fig:ctfidf]{Fig. \ref{fig:ctfidf}}). On the other hand, the long tail region beyond the Pareto-like distribution is composed of rare, often highly specialized terms that distinguish niche research contributions (e.g., “algae,” “engine,” and “probiotic”).

Together, these NLP results support the internal consistency and validity of the LLM-driven classification: high-frequency terms identified through BoW and c-TF-IDF models consistently aligned with the dominant themes discovered by the LLM. Moreover, the stark term concentration within specific topics, alongside adherence to known statistical patterns like Zipf's law and Pareto distributions, suggests that \textit{GPT-4o-mini} was not simply generating arbitrary groupings but instead uncovering meaningful and linguistically grounded structure.

\subsection{Alignment with PNAS Editorial Classifications}
To evaluate how the LLM-based classification captures cross-disciplinary patterns, the results were compared against the original PNAS dual-classification system \cite{Nat25}. PNAS frequently assigns research to two distinct scientific categories; this editorial “dual-labeling” serves as an external proxy for interdisciplinarity. Out of 1,493 validated research articles in the dataset, 661 were originally assigned to two PNAS categories. This yields a baseline dual-classification rate of 44.3\%, representing the average frequency of interdisciplinary labeling across the entire engineering collection.

Lift serves as a measure of cross-topic enrichment (or duality). A Lift value greater than 1.0 indicates that a topic is disproportionately composed of articles that the journal editors deemed cross-domain. Conversely, a Lift below 1.0 indicates a more “discipline-specific” or "siloed" topic where articles rarely bridge editorial categories. To illustrate these calculations, consider Topic 08 -- Synthetic Biology (\hyperref[tab:duality]{Table \ref{tab:duality}}). The LLM assigned 28 articles to this topic, 24 of which were dually classified by PNAS. The metrics are calculated as follows:
\begin{itemize}
\item Precision: $24 / 28 = 85.7\%$ (a majority of articles in this topic were dually classified).
\item Recall: $24 / 661 = 3.6\%$ (this topic accounts for a small fraction of all dually classified articles in the dataset).
\item Lift: $0.857 / 0.443 =  1.94$ (articles in this topic are nearly twice as likely to be dually classified compared to the average article in the dataset).
\end{itemize}

As shown in \hyperref[tab:duality]{Table \ref{tab:duality}}, biological-oriented engineering topics (e.g., Diagnostic Technologies, Tissue Engineering) consistently show high Lift, suggesting these fields are inherently integrative. In contrast, fundamental engineering topics like Material Science (0.19) and Soft Robotics (0.11) show significantly lower Lift. This does not imply a lack of complexity, but rather that these papers typically remain within the bounds of the single PNAS engineering category, reflecting a more specialized disciplinary focus.

\begin{table}[h]
\centering
\caption{Comparison of LLM-derived topics to the original PNAS dual classification system. "No. Abs" represents the number of abstracts classified in each topic, "Dual" indicates dual-classifications, and the metrics are calculated relative to the baseline dual classification.}
\begin{tabularx}{\textwidth}{@{\extracolsep{\fill}} lrrrrrl}
\toprule
Topic ($T_J$) & No. Abs & Dual & Prec. (\%) & Recall (\%) & Lift & Duality \\
\midrule
08 -- Synthetic Biology & 28 & 24 & 85.7 & 3.6 & 1.94 & Very High \\
05 -- Diagnostic Technologies & 202 & 162 & 80.2 & 24.5 & 1.81 & Very High \\
02 -- Nanoparticle Technology ... & 144 & 110 & 76.4 & 16.6 & 1.73 & Very High \\
01 -- Tissue Engineering ... & 253 & 189 & 74.7 & 28.6 & 1.69 & Very High \\
15 -- Bioengineering Innovations & 21 & 14 & 66.7 & 2.1 & 1.51 & Very High \\
\midrule
09 -- Bioinspired Robotics & 44 & 27 & 61.4 & 4.1 & 1.39 & High \\
\midrule
16 -- Imaging Techniques & 10 & 4 & 40.0 & 0.6 & 0.90 & Low \\
17 -- Other & 146 & 55 & 37.7 & 8.3 & 0.85 & Low \\
07 -- Biomedical Engineering & 58 & 21 & 36.2 & 3.2 & 0.82 & Low \\
13 -- Microfluidics Innovations & 25 & 7 & 28.0 & 1.1 & 0.63 & Low \\
\midrule
12 -- Water Purification & 25 & 3 & 12.0 & 0.5 & 0.27 & Very Low \\
04 -- Electronic Innovations & 108 & 12 & 11.1 & 1.8 & 0.25 & Very Low \\
12 -- Fluid Dynamics & 36 & 4 & 11.1 & 0.6 & 0.25 & Very Low \\
06 -- Material Science & 201 & 17 & 8.5 & 2.6 & 0.19 & Very Low \\
03 -- Catalysis and Energy & 122 & 10 & 8.2 & 1.5 & 0.19 & Very Low \\
10 -- Soft Robotics & 43 & 2 & 4.7 & 0.3 & 0.11 & Very Low \\
11 -- Shape-Morphing Materials & 27 & 0 & 0.0 & 0.0 & 0.00 & Very Low \\
\midrule
\textbf{Baseline} & \textbf{1,493} & \textbf{661} & \textbf{44.3} & \textbf{100.0} & \textbf{1.00} & $\sim$ \\
\bottomrule
\end{tabularx}
\label{tab:duality}
\end{table}

\section{Thematic Mapping of Full-Text}

\subsection{Secondary Full-Text Classification}
While abstract-level classification provided a useful summary of each article's primary focus, it may overlook secondary or embedded themes in the full text. 1,490 articles were successfully parsed, segmented, and classified, resulting in a dataset of 18,636 discrete texts. Unlike the primary classification, which assigned a single mutually exclusive topic per abstract ($y_i \in S^* | x_i$), the secondary pass allowed segments ($z_{ik} \in Z_i$) to be independently classified under one or more topics ($u_{ik} \subseteq S^* | z_{ik}$).

\begin{figure}[h]
\centering
\includegraphics[width=0.5\textwidth]{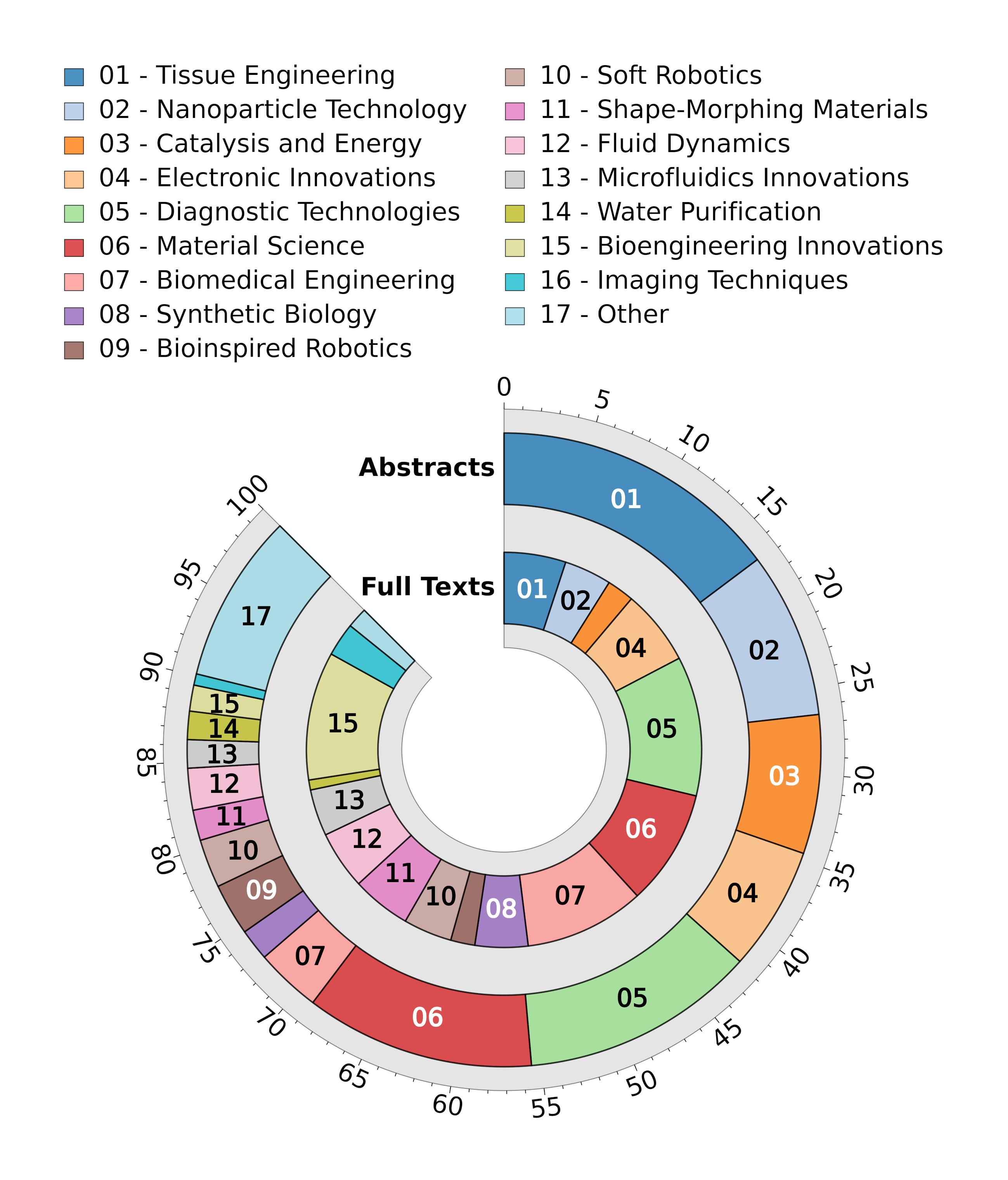}
\caption{Distribution of LLM-derived topic classifications for PNAS engineering. The outer ring ($n=1,519$) represents the single, primary classification of abstracts, while the inner ring ($n=18,636$) represents the multi-label classification of full-text segments.}
\label{fig:dist}
\end{figure}

\hyperref[fig:dist]{Fig. \ref{fig:dist}} compares these distributions. The relationship between the rings is comparative rather than hierarchical; both rings utilize the same 16 topic schema ($S^*$), including the $n^*+1$ “Other” category, but they represent different units of analysis (abstracts vs. segments). The outer ring shows the primary focus of the articles based on abstracts, while the inner ring captures the topics covered in full-text. The variance in proportions between rings (e.g., the expansion of Topic 15 in the inner ring) indicates that certain topics serve as broad foundations common across many segments, even if they are not the primary subject of the article's abstract.

The substantial shifts in proportions between the two rings, most notably the decrease in Topic 01 (Tissue Engineering) and the increase in Topic 15 (Bioengineering Innovations), reflect the transition from “application-domain” labeling to “technical-process” labeling.  For instance, while an article's primary identity (abstract) may be Tissue Engineering (Topic 01), the majority of its segments may describe the specific Bioengineering Innovations (Topic 15) or Diagnostic Technologies (Topic 05) used to achieve those results. Thus, Topic 01 acts as a dominant anchor category for abstracts, whereas Topic 15 serves as a cross-topic element that appears frequently across segments but is rarely the sole focus of an entire paper. \hyperref[tab:fulltext_dist]{Appendix, Table \ref{tab:fulltext_dist}} summarizes how each of these topics was expanded or consolidated based on secondary classification results.

Three distinct cases arise from this analysis: Case 1 -- Single, Specific Topic Assignment, Case 2 -- Indefinite “Other” Assignment, and Case 3 -- Multiple Topic Assignments. Of the segments analyzed, 25\% were classified into either a single dominant topic or exclusively as “Other”: 3,735 segments (Case 1, 20\%) showed clear topical focus, while 957 (Case 2, 5\%) could not be confidently assigned to any topic. The remaining 75\% were assigned multiple topics (Case 3), demonstrating the extensive breadth of the research and expanding the number of effective classifications to 46,639. On average, these multi-topic segments were linked to 2.6 ± 1.5 topics. Segments with more than four topic assignments were atypical and marked by ambiguity or overly general language, potentially prompting the model to “guess” its classification.

\subsection{Cross-Topic Relationships}
A bipartite graph was constructed to analyze relationships between abstract-level topics and the multi-label classifications derived from full-text segments. Each abstract topic ($y_i$) was connected to the topics assigned to its corresponding segments ($u_{ik} \subseteq S^*$), forming an adjacency matrix $M$ (\hyperref[tab:adjaceny_mat]{Appendix, Table \ref{tab:adjaceny_mat}}). Rows represent the primary abstract classification, while columns capture the distribution of segment-level classifications derived from the same articles.

Since segments were classified independently from abstracts, their topics frequently differed from the primary classification (previously described as “Case 3”). For example, the article “Brittle and ductile yielding in soft materials” \cite{Kam24} was classified as Material Science (Topic 06) at the abstract level, yet its 14 segments were distributed across Material Science, Soft Robotics, Fluid Dynamics, and Other, demonstrating a common cross-topic spread.

Row-wise analysis of $M$ reveals how full-text content distributes relative to the parent abstract category. For instance, among the 2,136 segments from articles classified as Material Science (row 6, Topic 06), 664 (31.1\%) were assigned exclusively to that topic (\hyperref[tab:fulltext_dist]{Appendix, Table \ref{tab:fulltext_dist}}) -- demonstrating “Case 1.” Due to multi-label tolerance, the segments are distributed as follows: 1,187 (40.4\%) preserved as Material Science ($u_{ik}=y_i$), and 2,686 (49.6\%) were situated in other subdomains ($u_{ik} \subseteq S^*$ where $u_{ik} \neq y_i$). Conversely, a column-wise summation of $M$ reveals that the Material Science category accumulated 3,561 segment labels from articles originally classified under other topics, totaling 5,378 labels or 10.8\% of the entire corpus.

\begin{figure}[h]
\centering
\includegraphics[width=0.90\textwidth]{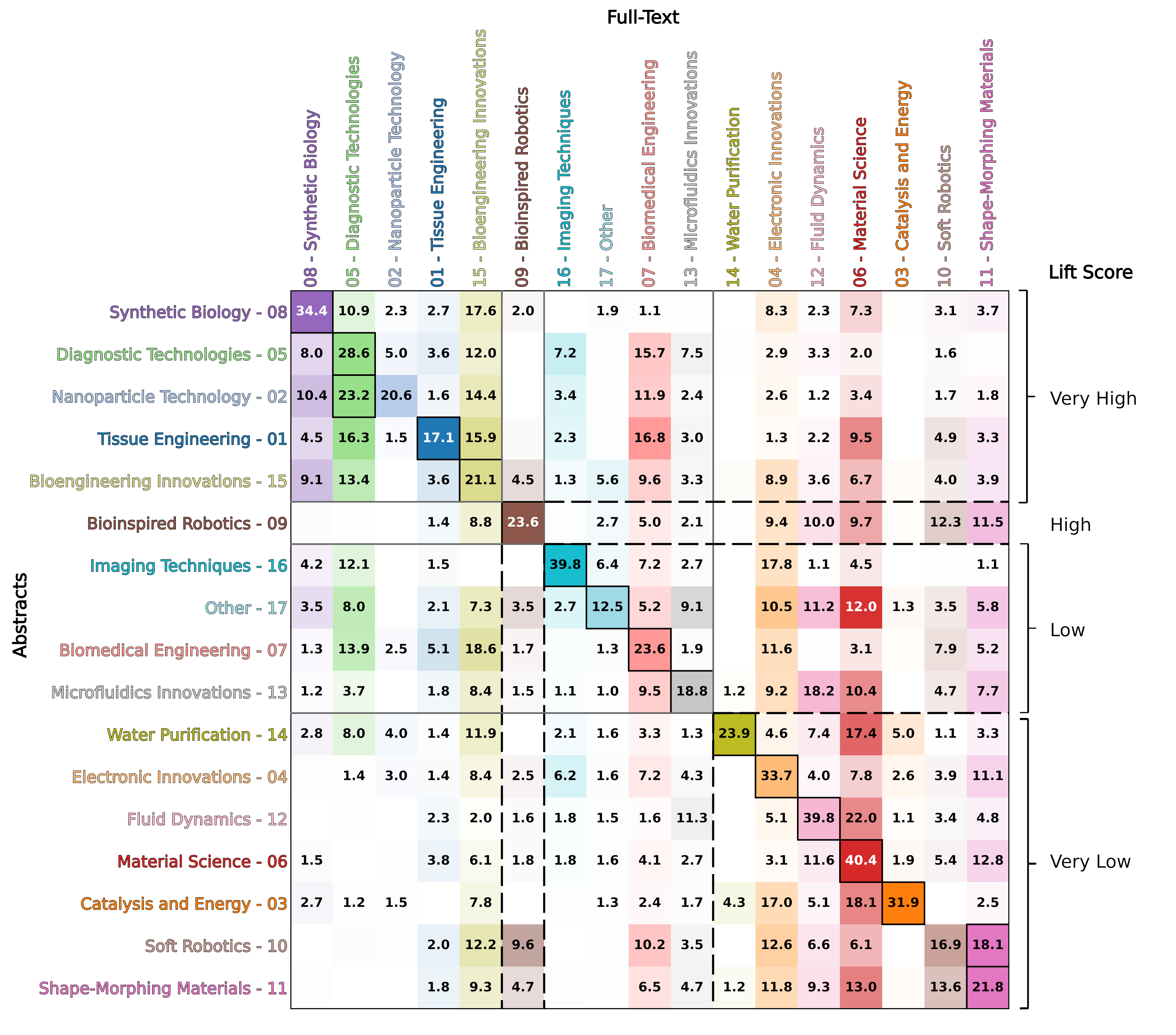}
\caption{Abstract (row) to full-text (column) normalized adjacency matrix $M$ of LLM-derived engineering topics. The index from “Very High” to “Very Low” indicates the degree of interdisciplinary from the lift score based on abstract classification relative to the original PNAS dual-classification system.}
\label{fig:matrix}
\end{figure}

The adjacency matrix is visualized as a normalized heatmap in \hyperref[fig:matrix]{Fig. \ref{fig:matrix}}, where rows represent a topic's full-text distribution scaled from 0 to 100\%. Connections below 1\% were omitted to highlight some inherent sparsity (e.g., Bioinspired Robotics having little representation in Water Purification, and vice versa). The diagonal represents strong topic self-alignment ($u_{ik}=y_i$), while off-diagonal concentrations reveal cross-topic relationships. Nanoparticle Technology (Topic 02) and Soft Robotics (Topic 10) were notable exceptions, slightly deviating from this trend. These off-diagonal patterns also reveal asymmetric dependencies between topics. For example, while only 1.9\% of Material Science (Topic 06) segments were co-classified as Catalysis and Energy (Topic 03), 18.1\% of segments from Catalysis and Energy were co-classified as Material Science, suggesting that Material Science often serves as a supporting discipline for other domains.

Finally, topics are ordered by descending lift values (\hyperref[tab:duality]{Table \ref{tab:duality}}) to compare LLM-derived classifications with known dual-classification indicators. Topics with “Very High” lift values (e.g., Synthetic Biology) frequently connect with other life-science themes, while engineering-dominant topics with “Very Low” lift values (e.g., Material Science and Fluid Dynamics) largely connect with each other. In contrast, the “High” and “Low” lift groups, containing more mixed-domain topics like Bioinspired Robots, exhibit a bridging behavior. Their full-text segments often connect biologically oriented abstracts to engineering methods.

\section{Discussion}

\subsection{Abstract Topic Modeling with LLMs}
The proposed abstract classification method was refined over $T=9$ iterations, resulting in $n^*=16$ stable categories, during which topic descriptions ($T_j^t, D_j^t$) were revised until a convergence threshold ($\delta=0.10$) was satisfied. Articles that consistently failed to align with any stable topic were assigned to an additional “Other” category ($n^* + 1$). This iterative refinement allowed topic descriptions to gradually stabilize while preventing poorly defined clusters from absorbing unrelated articles.

The evolution of water-related topics illustrates this process. In an early iteration ($t=2$), a broad “Water Sustainability” topic emerged from 132 abstracts. However, only 16.7\% of those abstracts aligned with the generated topic description, which emphasized governance strategies for water purification. After several refinement steps, the cluster evolved into the more focused topic “Water Purification.” The revised description, emphasizing membrane technologies and wastewater treatment, increased the agreement score ($A_j^t$) to 66.7\% (Topic 14, \hyperref[tab:cluster_agreement]{Appendix, Table \ref{tab:cluster_agreement}}). This demonstrates that the method can be adapted to
different research areas, ensuring higher topic agreement and stability with a more focused definition when applied to a smaller, more relevant subset of abstracts.

The effectiveness of this framework was supported by comparisons with other topic modeling methods. As shown in \hyperref[tab:topicmodels]{Table \ref{tab:topicmodels}}, the proposed method achieved the highest topic diversity ($T_D=0.909$), the lowest topic overlap (Jaccard $=6.15\times10^{-3}$), and competitive coherence scores ($C_V=0.446$, UMass $=-6.42$) among the evaluated approaches. While MALLET LDA achieved the strongest coherence scores, its lower diversity and higher similarity values suggest greater overlap between topics. Moreover, the LLM-based approaches demonstrate similar performance trends but reflect different methodological assumptions. In the TopicGPT framework, topics are discovered by prompting the LLM to freely propose candidate topics for individual documents, followed by a consolidation step \cite{pham_topicgpt_2024} -- which produced 19 topics. This model yielded diversity and similarity scores close to those of the proposed method; however, the absence of outlier handling can lead to forced topic assignment.

The PDDP approach, in contrast, employs a hierarchical clustering strategy guided by LLM-assisted interpretation \cite{theo_2025}. For the same number of topics ($n^*=16$), this model achieved comparable evaluation metrics but also omits outlier handling. Even BERTopic \cite{Gro22}, which supports outlier detection, exhibited higher topic similarity and lower diversity. Thus, the proposed framework differs from these methods by combining iterative topic refinement with explicit convergence criteria ($\tau$, $\delta$) and a dedicated “Other” category for poorly aligned articles. This design allows topic definitions to stabilize over multiple iterations while preventing weakly related documents from being assigned to an unrelated topic. 

While automated metrics are insightful, they should not necessarily be interpreted as definitive indicators of topic quality, especially in the absence of ground-truth labels \cite{wu_survey_2024}. Therefore, a manual validation was conducted over a random sample of 303 abstracts (20\% of the corpus) to further assess the reliability of the LLM-based classification. When evaluated using strict single-label agreement, the LLM classification achieved a Top-1 accuracy of 52.5\%. However, allowing multiple conceptually valid themes increased the Top-3 accuracy to 75.9\% (\hyperref[tab:manual]{Table \ref{tab:manual}}). This result reflects the cross-topic nature of the corpus, where many articles span multiple engineering subdomains. These insights were further expanded through comprehensive NLP analyses and secondary classification results, as detailed in the subsequent discussion.

\subsection{Limitations of Clustering Algorithms}
The findings from this study also highlight the inherent limitations of standard clustering algorithms like K-means for scientific texts. Finding an optimal number of clusters ($k$) is a well-known challenge. Common heuristics such as the elbow method and Silhouette score were unreliable for this dataset and showed no clear optimum (\hyperref[fig:nclusters]{Appendix, Fig. \ref{fig:nclusters}}), likely due to data sparsity and outliers \cite{Shi21}. While other algorithms like DBSCAN are designed to identify outliers, their performance can be hindered in high-dimensional embedding spaces \cite{Dua21}. This study also assumes that an appropriate $k$ can be predefined or selected heuristically for the data set at each recursion step. Future implementations should explore adaptive methods for dynamic $k$ selection or perform sensitivity testing with other clustering algorithms.

A further limitation of clustering is the lack of direct convergence criteria. The proposed approach addresses this by incorporating user-defined thresholds ($\tau$ and $\delta$) to govern the topic stabilization process. Without such controls, the number of topics may not converge, creating overly specialized categories with insufficient representation. For instance, in a late iteration ($T=9$), a potential “Imaging Techniques” topic included only ten abstracts. Allowing such small topics to form would be analogous to arbitrarily increasing the cluster count in a standard algorithm, undermining the goal of identifying significant research areas. 

The LLM-based reclassification and agreement criteria provided a practical and robust framework for identifying context-driven research topics. A primary advantage is its direct interpretability; generating descriptive labels and descriptions for each cluster immediately clarifies the substance of topics. Furthermore, this iterative method offers an important technical benefit. By processing a smaller, controlled number of clusters ($k$) in each iteration, the framework mitigates the “lost in the middle” problem, where LLM performance degrades when processing large contexts \cite{Liu24}. This approach reduces the input complexity for the LLM at each step, ensuring more predictable and accurate classification.

\subsection{NLP Insights on LLM-Generated Topics}
The stable topics identified through primary abstract classification were further interpreted using established NLP methods. Analyses with BoW and c-TF-IDF confirmed that the sixteen generated topics were not arbitrary; rather, each reflected a distinct vocabulary and research focus. For example, clear differences in term distributions between Tissue Engineering (Topic 01) and Material Science (Topic 06) support the model's ability to distinguish disciplinary boundaries.

The word frequency distribution of the corpus broadly followed Zipf's law \cite{Zip36}. Although the empirical distribution did not perfectly match the theoretical curve (\hyperref[fig:zipf]{Fig. \ref{fig:zipf}b}), the data exhibited the characteristic structure in which a small number of terms occur frequently (e.g., “study,” “method”), while a long tail of terms appears rarely (e.g., “mesenchymal,” “photovoltaic”). This linguistic pattern highlights the advantage of LLMs, which leverage attention mechanisms to weigh both frequent and rare terms within context \cite{Vas17}: the term “dislocation” appeared almost exclusively within the Material Science topic in PNAS, likely referring to atomic-scale defects rather than joint dislocations (e.g., shoulder, knee) in biomedical contexts.

\textit{GPT-4o-mini}'s contextual sensitivity was also evident in its handling of low-frequency but domain-specific terminology. For instance, “microfluidics” appeared frequently as an author keyword (\hyperref[fig:ctfidf]{Fig. \ref{fig:ctfidf}}), but ranked relatively low in overall word frequency ($f=17$, $r=587$). Despite this, the primary classification approach identified its importance by generating a dedicated Microfluidics Innovation topic (Topic 13), which expanded from 1.6\% of abstracts to 4.4\% of the full-text corpus. Similarly, the model resolved contextual ambiguity in multi-word expressions: while the bigram “mechanical properties” was strongly associated with Material Science, the term “mechanical” in Tissue Engineering contexts more often referred to physical stimuli applied to cells. The LLM's capacity to process intermittent multi-word phrases was critical for accurate topic separation.

The absence of a dedicated topic for “machine learning” or “deep learning,” despite their prominence as author keywords from 2020 to 2024 (\hyperref[fig:ctfidf]{Fig. \ref{fig:ctfidf}}), was also an informative finding. This pattern likely reflects the structure of the original PNAS classification system, in which Computer Science represents only 0.38\% of all published articles and none of the Engineering articles were dually classified under that category (\hyperref[tab:pnas_pubcount]{Appendix, Table \ref{tab:pnas_pubcount}}). As AI-driven methods have become integral to modern research, these topics may seek more representation in the future.

While frequency-based approaches such as BoW and c-TF-IDF provide functional model interpretation and confirmation, they remain inherently context-agnostic. These models treat each word in isolation, unable to infer meaning beyond occurrence counts. In contrast, the LLM-based approach successfully identified latent semantic structures within the corpus. This was reflected in the alignment between high-frequency BoW terms (e.g., “cell,” tissue,” “matrix”) and the dominant topic identified by the model (Tissue Engineering). Such consistency between statistical patterns and semantic clustering suggests that the LLM uncovered a meaningful, linguistically grounded structure in the text, highlighting its strength for exploratory analysis where no formal ground-truth is available.

\subsection{Cross-Topic Connections in Engineering Research}
While the primary analyses evaluated the quality and coherence of the generated topics, it is equally important to examine how these topics interact across the corpus. Contrary to criticisms that PNAS underrepresents engineering \cite{milojević_nature-science_2020}, the secondary, full-text analysis reveals its strong presence within adjacent domains through cross-topic relationships. Connections between abstract-level topics and their corresponding full-text segments were represented as a bipartite graph using an asymmetric adjacency matrix (\hyperref[tab:adjaceny_mat]{Appendix, Table \ref{tab:adjaceny_mat}}). This structure revealed a complex network of relationships that mirrored and expanded upon the insights from the journal's original dual-classification system.

To demonstrate, the LLM-generated category of Tissue Engineering emerged as a highly cross-topic field relative to the PNAS dual-classification proxy (\hyperref[tab:duality]{Table \ref{tab:duality}}), achieving high precision (73.8\%) and recall (28.6\%). The secondary classification analysis independently reinforced this finding, positioning Tissue Engineering as a central hub connecting multiple research domains. This qualitative observation is further supported by corpus-level statistics: while the journal's baseline dual-classification rate for engineering articles is 44.3\%, the full-text analysis assigned a comparable proportion of segments to multiple topics (42.5\%, 21,074 segments) associated with high-duality categories (“Very High” and “High” lift). The remaining 57.5\% of segments were concentrated in more engineering-centered topics that are less frequently dually classified under the original PNAS taxonomy.

The normalized adjacency matrix (\hyperref[fig:matrix]{Fig. \ref{fig:matrix}}) provides further insights into the journal's topical research structure. Strong diagonal values, where segments align with their parent abstract's topic, confirm the thematic coherence of the classifications. However, several topics exhibited notable cross-topic activity. For instance, Nanoparticle Technology (Topic 02) and Soft Robotics (Topic 10) frequently appeared as secondary themes embedded within broader biomedical narratives. Segments from Nanoparticle Technology abstracts were often classified under Diagnostic Technologies, reflecting applications such as the T-cell isolation nano-vials developed by Koo et al. \cite{Koo24}. The matrix also revealed emerging areas not captured by the journal's formal taxonomy. Shape-Morphing Materials, for example, exhibited substantial segment activity despite having no dual classifications in the original PNAS system, suggesting a growing but previously obscured research direction.

Finally, the analysis highlights the foundational role of dominant engineering topics such as Material Science (Topic 06) and Fluid Dynamics (Topic 12). These topics displayed strong self-correlation while maintaining significant connections to other areas, indicating they function as technical foundations for many bio-oriented innovations. Their prevalence reinforces the convergence, rather than mere coexistence, of biological and engineering domains. Together, these patterns affirm that engineering within PNAS operates through deep interdisciplinary integration and that the LLM-driven framework not only corroborates the original classification system but expands upon it, offering a dynamic and adaptable method for mapping cross-topic relationships in modern scientific research.

\subsection{Challenges with LLMs}
Despite impressive advances, LLMs have limitations, particularly the risk of hallucination where a model generates plausible-sounding but incorrect or fabricated information \cite{Hua23,Far24}. These risks were intentionally engaged as this study employed an exploratory, unsupervised classification system. Nevertheless, the results demonstrated the robustness and potential of LLMs for scientific text mining, instilling optimism about the future of their application.

Several structured safeguards must be implemented to mitigate the risks of bias and hallucination. To combat this, the order of abstracts was shuffled before LLM classification to reduce sequence-dependent bias, which can influence predictions in autoregressive models when inputs are batched (e.g., more than one text is labeled simultaneously). Additionally, a “3-out-of-5” consensus rule was enforced, meaning an abstract had to be consistently assigned to the same topic in at least three out of five runs to be retained for subsequent analysis. The high consistency of the model was evidenced by the fact that this criterion filtered out only 5\% of abstracts (\hyperref[fig:3of5_criteria]{Appendix, Fig. \ref{fig:3of5_criteria}}), reinforcing the reliability of the classification process.

In future studies, orchestrating multiple LLMs across different architectures or model providers could only enhance classification. Additional improvements may also come from more advanced text parsing and extraction techniques, especially for handling domain-specific artifacts such as complex chemical names or specialized terminology in scientific publications. While these models are not without flaws, their ability to infer complex structures from highly technical jargon, especially without additional supervised training, demonstrates their promising role in exploratory research and trend detection across disciplines.

Overall, classical NLP techniques remained indispensable for validating \textit{GPT-4o-mini}'s outputs. While LLMs can operate as “black-box” systems, interpretable models such as BoW and c-TF-IDF allow for the validation and contextualization of patterns they detect. In this study, these traditional tools were not merely complementary but essential for confirming the semantic coherence of the LLM-assigned topics. This hybrid approach verified that the LLM-derived groupings aligned with known linguistic structures and underscored the value of blending modern generative AI capabilities with established, interpretable NLP techniques.

\subsection{Future Analyses of the Engineering Coverage in PNAS}
While this study presents a stable snapshot of engineering research in PNAS, the true value of the proposed methodology lies in its dynamic application for future analysis. The sixteen topics identified provide a robust baseline of the current landscape. However, as scientific frontiers evolve, this classification framework is designed to adapt. New research trends will inevitably emerge, similar to how topics like Catalysis and Energy (Topic 03) have gained prominence in recent years (\hyperref[fig:dist1]{Fig. \ref{fig:dist1}}). Such a topic may not have surfaced as a distinct cluster from the abstracts prior to 2015, demonstrating how this approach can track the rise and fall of research paradigms over time.

The current classifications can be preserved as a benchmark for the immediate future. This provides a valuable resource for prospective authors, allowing them to situate their work within the context of related engineering research previously published in PNAS. The stability of these themes, as confirmed by the NLP validation, offers confidence that they represent significant and persistent areas of inquiry. Based on the analysis, a reasonable prediction is that the strong influence of biologically oriented research on these topics will continue, ensuring that many of the current themes remain relevant in the journal.

The long-term utility of this framework is its capacity for simple re-evaluation. As new research develops, for instance, at the intersection of Engineering and Computer Science, the “Other” category will likely be ever-changing and fluid. A significant increase in the size of this category can serve as a trigger to re-run the classification algorithm. This forward analysis would allow for identifying new, emerging subtopics and could reveal shifts in the relevance of existing ones. For example, specific topics may expand, while others, which were once considered “cutting edge,” may shrink or merge as the field's focus evolves. This dynamic reclassification is a key feature of the methodology, enabling a continuous and updated understanding of the topical landscape -- unless the “Other” category remains stable.

\section{Conclusion}
As a well-known and impactful scientific journal, the \textit{Proceedings of the National Academy of Sciences} (PNAS) plays a central role in shaping modern scientific discourse. Understanding the composition of its published work is therefore essential for researchers and the public alike. This study demonstrates how an LLM-driven topic modeling framework can transform the mapping of scientific literature by revealing both explicit thematic identities and latent cross-topic structures within a 20-year engineering corpus. By combining data-driven unsupervised learning with the contextual intelligence of LLMs, the methodology translates dense, specialized jargon into a shared, interpretable thematic space. This approach moves beyond the limitations of traditional “bag-of-words” models, which often require intensive domain expertise to interpret.

The proposed pipeline combines abstract-level topic identification with a secondary full-text classification stage, allowing both primary themes and embedded secondary topics to be detected. The article abstracts were grouped into sixteen major topics (e.g., Material Science, Tissue Engineering) through a recursive K-means clustering and LLM classification algorithm. The reliability of this framework was confirmed through comparative evaluation with established topic modeling methods, achieving 27\% higher topic diversity and 61\% lower topic overlap, on average, compared to other models as well as competitive $C_V$ and UMass coherence scores. Moreover, the implementation of a persistent “Other” category (representing 9.8\% of abstracts) acts as a real-time trigger for model re-evaluation. This allows the framework to detect emerging subtopics and “noisy” data that traditional models often conflate, ensuring the thematic structure remains high-quality as scientific fields evolve. 

A manual review of a randomly sampled subset of abstracts showed substantial agreement between the model and author annotations. This review resulted in an accuracy of 52.5\% when only one top-ranking human labeled topic is valid and 75.9\% when three are considered. These accuracies reflect the inherent cross-topic nature of the corpus, where many articles span multiple engineering subdomains, which is explored in greater detail in the secondary full-text analysis. A bipartite graph between abstract-level topics and segment-level classifications exposes thematic connections that are not easily captured through abstract or keyword analyses alone. Moreover, the graph structure broadly aligns with the journal's editorial dual-classification system while also revealing additional latent topic interactions, such as the frequent intersections between engineering and adjacent domains like biology or medicine.

Identifying such cross-topic trends is critical for researchers tracking their fields' trajectory and the wider community, as scientific literature is often dense and difficult to comprehend. This work highlights the potential of LLMs to address this challenge. The selected model demonstrated an ability to generate concise, plain-language descriptions for highly technical research areas, translating dense jargon into accessible topics. This capability has significant implications for education and interdisciplinary collaboration, helping to bridge the gap between experts and broader audiences. Looking ahead, further advancements in LLMs, such as improved language understanding and generation capabilities, could significantly enhance their utility in scientific communication and understanding.

These results suggest that LLM-based topic modeling can provide a robust exploratory framework in domains where formal ground-truth labels are unavailable. While these results are promising, the study also acknowledges the limitations of LLMs, particularly the risk of hallucination -- a discrepancy that is expected to improve with future advancements. However, the high consistency of classifications across multiple runs and the alignment with established NLP patterns suggest that the model's outputs are reliable for exploratory analysis. Furthermore, while the parameters of the proposed method (e.g., $\tau$, $\delta$, $k$) were selected based on empirical testing, future work should explore systematic approaches for optimizing these parameters to further enhance model performance.

Ultimately, this study provides a foundation for more dynamic, real-time mapping of scientific knowledge, highlighting thepotential of LLMs as powerful tools for large-scale literature analysis. By integrating contextual language understanding with traditional NLP validation techniques, the proposed framework offers a flexible approach for exploring topics in scientific corpora. As LLM capabilities continue to improve, such methods may play an increasingly important role in supporting literature synthesis, research discovery, and the automated mapping of science.

\bibliographystyle{unsrtnat}
\bibliography{references}

@book{Zip36,
  author    = {Zipf, George},
  title     = {{The Psycho-Biology of Language: an Introduction to Dynamic Philology}},
  publisher = {Houghton Mifflin},
  address   = {Boston, MA},
  year      = {1935},
  note      = {Reprinted by Routledge, 2014}
}

@article{zhang_scientific-large_2025,
  author  = {Zhang, Qiang and Ding, Keyan and Lv, Tianwen and others},
  title   = {{Scientific Large Language Models: A Survey on Biological \& Chemical Domains}},
  number  = {6},
  pages   = {1--38},
  note    = {doi \href{https://doi.org/10.1145/3715318}{10.1145/3715318}},
  volume  = {57},
  journal = {ACM Computing Surveys},
  year    = {2025}
}

@article{zawacki-richter_mapping_2016,
  author  = {Zawacki-Richter, Olaf and Naidu, Som},
  title   = {{Mapping research trends from 35 years of publications in Distance Education}},
  number  = {3},
  pages   = {245--269},
  note    = {doi: \href{https://doi.org/10.1080/01587919.2016.1185079}{10.1080/01587919.2016.1185079}},
  volume  = {37},
  journal = {Distance Education},
  year    = {2016}
}

@article{xie_feature-analysis_2018,
  author  = {Xie, Zheng and Li, Miao and Li, Jianping and Duan, Xiaojun and Ouyang, Zhenzheng},
  title   = {{Feature analysis of multidisciplinary scientific collaboration patterns based on PNAS}},
  pages   = {1--17},
  note    = {doi: \href{https://doi.org/10.1140/epjds/s13688-018-0134-z}{10.1140/epjds/s13688-018-0134-z}},
  volume  = {7},
  journal = {EPJ Data Science},
  year    = {2018}
}

@article{wallach_reproducible-research_2018,
  author  = {Wallach, Joshua and Boyack, Kevin and Ioannidis, John},
  title   = {{Reproducible research practices, transparency, and open access data in the biomedical literature, 2015-2017}},
  number  = {11},
  pages   = {e2006930},
  note    = {doi: \href{https://doi.org/10.1371/journal.pbio.2006930}{10.1371/journal.pbio.2006930}},
  volume  = {16},
  journal = {PLoS Biology},
  year    = {2018}
}

@article{verma_impact-not_2015,
  author  = {Verma, Inder},
  title   = {{Impact, not impact factor}},
  number  = {26},
  pages   = {7875--7876},
  note    = {doi: \href{https://www.pnas.org/doi/10.1073/pnas.1509912112}{10.1073/pnas.1509912112}},
  volume  = {112},
  journal = {Proceedings of the National Academy of Sciences},
  year    = {2015}
}

@inproceedings{Vas17,
  author       = {Vaswani, Ashish and Shazeer, Noam and Parmar, Niki and others},
  title        = {{Attention Is All You Need}},
  booktitle    = {Proceedings of the 31st Conference on Neural Information Processing Systems, NIPS},
  url          = {https://proceedings.neurips.cc/paper_files/paper/2017/file/3f5ee243547dee91fbd053c1c4a845aa-Paper.pdf},
  volume       = {30},
  organization = {Advances in Neural Information Processing Systems},
  address      = {Long Beach, California},
  publisher    = {Curran Associates, Inc.},
  year         = {2017}
}

@article{thapa_chatgpt-bard_2023,
  author  = {Thapa, Surendrabikram and Adhikari, Surabhi},
  title   = {{ChatGPT, Bard, and Large Language Models for Biomedical Research: Opportunities and Pitfalls}},
  pages   = {2647--2651},
  note    = {doi: \href{https://doi.org/s10439-023-03284-0}{10.1007/s10439-023-03284-0}},
  volume  = {51},
  journal = {Annals of Biomedical Engineering},
  year    = {2023}
}

@article{shyr_leveraging-artificial_2024,
  author  = {Shyr, Cathy and Grout, Randall and Kennedy, Nan and others},
  title   = {{Leveraging artificial intelligence to summarize abstracts in lay language for increasing research accessibility and transparency}},
  number  = {10},
  pages   = {2294--2303},
  note    = {doi: \href{https://doi.org/10.1093/jamia/ocae186}{10.1093/jamia/ocae186}},
  volume  = {31},
  journal = {Journal of the American Medical Informatics Association},
  year    = {2024}
}

@article{shiffrin_mapping-knowledge_2004,
  author  = {Shiffrin, Richard and Börner, Katy},
  title   = {{Mapping knowledge domains}},
  pages   = {5183--5185},
  note    = {doi: \href{https://www.pnas.org/doi/abs/10.1073/pnas.0307852100}{10.1073/pnas.0307852100}},
  volume  = {101},
  journal = {Proceedings of the National Academy of Sciences},
  year    = {2004}
}

@article{Shi21,
  author  = {Shi, Congming and Wei, Bingtao and Wei, Shoulin and Wang, Wen and Liu, Jialei},
  title   = {{A quantitative discriminant method of elbow point for the optimal number of clusters in clustering algorithm}},
  note    = {doi: \href{https://doi.org/10.1186/s13638-021-01910-w}{10.1186/s13638-021-01910-w}},
  volume  = {31},
  journal = {EURASIP Journal on Wireless Communications and Networking},
  year    = {2021}
}

@article{sharon_measuring-mumbo_2014,
  author  = {Sharon, Aviv and Baram-Tsabari, Ayelet},
  title   = {{Measuring mumbo jumbo: A preliminary quantification of the use of jargon in science communication}},
  note    = {doi: \href{https://doi.org/10.1177/0963662512469916}{10.1177/0963662512469916}},
  number  = {5},
  pages   = {528--546},
  volume  = {23},
  journal = {Public Understanding of Science},
  year    = {2014}
}

@article{schekman_charting-the_2008,
  author  = {Schekman, Randy},
  title   = {{Charting the course for PNAS}},
  number  = {8},
  pages   = {2755--2756},
  note    = {doi: \href{https://doi.org/10.1073/pnas.0800528105}{10.1073/pnas.0800528105}},
  volume  = {105},
  journal = {Proceedings of the National Academy of Sciences},
  year    = {2008}
}

@article{Pia14,
  author  = {Piantadosi, Steven},
  title   = {{Zipf's word frequency law in natural language: A critical review and future directions}},
  pages   = {1112--1130},
  note    = {doi: \href{https://doi.org/10.3758/s13423-014-0585-6}{10.3758/s13423-014-0585-6}},
  volume  = {21},
  journal = {Psychonomic Bulletin \& Review},
  year    = {2014}
}

@article{Pez18,
  author  = {Pezzoti, Nicola and Fekete, Jean-Daniel and Höllt, Thomas and others},
  title   = {{Multiscale Visualization and Exploration of Large Bipartite Graphs}},
  note    = {doi: \href{https://doi.org/10.1111/cgf.13441}{10.1111/cgf.13441}},
  number  = {3},
  pages   = {549--560},
  volume  = {37},
  journal = {Computer Graphics Forum},
  year    = {2018}
}

@article{petroșanu_tracing-the_2023,
  author  = {Petroșanu, Dana-Mihaela and Pîrjan, Alexandru and Tăbușcă, Alexandru},
  title   = {{Tracing the Influence of Large Language Models across the Most Impactful Scientific Works}},
  number  = {24},
  pages   = {4957},
  note    = {doi: \href{https://doi.org/10.3390/electronics12244957}{10.3390/electronics12244957}},
  volume  = {12},
  journal = {Electronics},
  year    = {2023}
}

@article{peng_a-study_2023,
  author  = {Peng, Cheng and Yang, Xi and Chen, Aokun and others},
  title   = {{A study of generative large language model for medical research and healthcare}},
  pages   = {1--17},
  note    = {doi: \href{https://doi.org/10.1038/s41746-023-00958-w}{10.1038/s41746-023-00958-w}},
  volume  = {6},
  journal = {npj Digital Medicine},
  year    = {2023}
}

@article{New05,
  author  = {Newman, Mark},
  title   = {{Power laws, Pareto distributions, and Zipf's law}},
  doi     = {10.1080/00107510500052444},
  number  = {5},
  pages   = {323--351},
  note    = {doi: \href{https://doi.org/10.1080/00107510500052444}{10.1080/00107510500052444}},
  volume  = {46},
  journal = {Contemporary Physics},
  year    = {2005}
}

@article{milojević_nature-science_2020,
  author  = {Milojević, Staša},
  title   = {{Nature, Science, and PNAS: disciplinary profiles and impact}},
  number  = {3},
  pages   = {1301--1315},
  note    = {doi: \href{https://doi.org/10.1007/s11192-020-03441-5}{10.1007/s11192-020-03441-5}},
  volume  = {123},
  journal = {Scientometrics},
  year    = {2020}
}

@article{markowitz_from-complexity_2024,
  author  = {Markowitz, David},
  title   = {{From complexity to clarity: How AI enhances perceptions of scientists and the public's understanding of science}},
  note    = {doi: \href{https://doi.org/10.1093/pnasnexus/pgae387}{10.1093/pnasnexus/pgae387}},
  number  = {9},
  pages   = {pgae387},
  volume  = {3},
  journal = {PNAS Nexus},
  year    = {2024}
}

@article{Man04,
  author  = {Mane, Ketan and Börner, Katy},
  title   = {{Mapping topics and topic bursts in PNAS}},
  pages   = {5287--5290},
  note    = {doi: \href{https://doi.org/10.1073/pnas.0307626100}{10.1073/pnas.0307626100}},
  volume  = {101},
  journal = {Proceedings of the National Academy of Sciences},
  year    = {2004}
}

@article{luzón_public-communication_2013,
  author  = {Luzón, María},
  title   = {{Public Communication of Science in Blogs: Recontextualizing Scientific Discourse for a Diversified Audience}},
  note    = {doi: \href{https://doi.org/10.1177/0741088313493610}{10.1177/0741088313493610}},
  number  = {4},
  pages   = {428--457},
  volume  = {30},
  journal = {Written Communication},
  year    = {2013}
}

@article{lupia_trends-in_2024,
  author  = {Lupia, Arthur and Allison, David and Jamieson, Kathleen and others},
  title   = {{Trends in US public confidence in science and opportunities for progress}},
  number  = {11},
  pages   = {e2319488121},
  note    = {doi: \href{https://doi.org/10.1073/pnas.2319488121}{10.1073/pnas.2319488121}},
  volume  = {121},
  journal = {Proceedings of the National Academy of Sciences},
  year    = {2024}
}

@article{Liu24,
  author  = {Liu, Nelson and Lin, Kevin and Hewitt, John and others},
  title   = {{Lost in the Middle: How Language Models Use Long Contexts}},
  pages   = {157--173},
  note    = {doi: \href{https://doi.org/10.1162/tacl_a_00638}{10.1162/tacl\_a\_00638}},
  volume  = {12},
  journal = {Transactions of the Association for Computational Linguistics},
  year    = {2024}
}

@article{Koo24,
  author  = {Koo, Doyeon and Mao, Zhiyuan and Noguchi, Mikayo and others},
  title   = {{Defining T cell receptor repertoires using nanovial-based binding and functional screening}},
  note    = {doi: \href{https://doi.org/10.1073/pnas.2320442121}{10.1073/pnas.2320442121}},
  number  = {14},
  pages   = {e2320442121},
  volume  = {121},
  journal = {Proceedings of the National Academy of Sciences},
  year    = {2024}
}

@article{Kam24,
  author  = {Kamani, Krutarth and Rogers, Simon},
  title   = {{Brittle and ductile yielding in soft materials}},
  number  = {22},
  pages   = {e2401409121},
  note    = {doi: \href{https://doi.org/10.1073/pnas.2401409121}{10.1073/pnas.2401409121}},
  volume  = {121},
  journal = {Proceedings of the National Academy of Sciences},
  year    = {2024}
}

@article{ioannidis_meta-research_2018,
  author  = {Ioannidis, John},
  title   = {{Meta-research: Why research on research matters}},
  note    = {doi: \href{https://doi.org/10.1371/journal.pbio.2005468}{10.1371/journal.pbio.2005468}},
  pages   = {e2005468},
  number  = {3},
  volume  = {16},
  journal = {PLoS Biology},
  year    = {2018}
}

@article{Hua23,
  author  = {Huang, Jie and Chang, Kevin},
  title   = {{Towards Reasoning in Large Language Models: A Survey}},
  doi     = {10.48550/arXiv.2212.10403},
  note    = {doi: \href{https://doi.org/10.48550/arXiv.2212.10403}{10.48550/arXiv.2212.10403}},
  journal = {arXiv preprint arXiv:2212.10403},
  year    = {2023}
}

@article{guo_personalized-jargon_2024,
  author  = {Guo, Yue and Chang, Joseph and Antoniak, Maria and others},
  title   = {{Personalized Jargon Identification for Enhanced Interdisciplinary Communication}},
  note    = {doi: \href{https://doi.org/10.48550/arXiv.2311.09481}{10.48550/arXiv.2311.09481}},
  journal = {arXiv preprint arXiv:2311.09481},
  year    = {2024}
}

@article{guo_automated-lay_2021,
  author  = {Guo, Yue and Qiu, Wei and Wang, Yizhong and Cohen, Trevor},
  title   = {{Automated Lay Language Summarization of Biomedical Scientific Reviews}},
  note    = {doi: \href{https://doi.org/10.48550/arXiv.2012.12573}{10.48550/arXiv.2012.12573}},
  journal = {arXiv preprint arXiv:2012.12573},
  year    = {2020}
}

@article{Gro22,
  author  = {Grootendorst, Marten},
  title   = {BERTopic: Neural topic modeling with a class-based TF-IDF procedure},
  note    = {doi: \href{https://doi.org/10.48550/arXiv.2203.05794}{10.48550/arXiv.2203.05794}},
  journal = {arXiv preprint arXiv:2203.05794},
  year    = {2022}
}

@article{field_public-understanding_2001,
  author  = {Field, Hyman and Powell, Patricia},
  title   = {{Public understanding of science versus public understanding of research}},
  note    = {doi: \href{https://journals.sagepub.com/doi/10.3109/a036879}{10.3109/a036879}},
  number  = {4},
  pages   = {421--426},
  volume  = {10},
  journal = {Public Understanding of Science},
  year    = {2001}
}

@article{Far24,
  author  = {Farquhar, Sebastian and Kossen, Jannik and Kuhn, Lorenz and Gal, Yarin},
  title   = {{Detecting hallucinations in large language models using semantic entropy}},
  pages   = {625--630},
  note    = {doi: \href{https://doi.org/10.1038/s41586-024-07421-0}{10.1038/s41586-024-07421-0}},
  volume  = {630},
  journal = {Nature},
  year    = {2024}
}

@article{Dua21,
  author  = {Duan, Liang and Ma, Shuai and Aggarwal, Charu and Saket, Sather},
  title   = {{Improving spectral clustering with deep embedding, cluster estimation and metric learning}},
  pages   = {675--694},
  note    = {doi: \href{https://doi.org/10.1007/s10115-020-01530-8}{10.1007/s10115-020-01530-8}},
  volume  = {63},
  journal = {Knowledge and Information Systems},
  year    = {2021}
}

@article{ding_disciplinary-structures_2018,
  author  = {Ding, Jielan and Ahlgren, Per and Yang, Liying and Yue, Ting},
  title   = {{Disciplinary structures in Nature, Science and PNAS: journal and country levels}},
  number  = {3},
  pages   = {1817--1852},
  note    = {doi: \href{http://doi.org/10.1007/s11192-018-2812-9}{10.1007/s11192-018-2812-9}},
  volume  = {116},
  journal = {Scientometrics},
  year    = {2018}
}

@article{dagdelen_structured-information_2024,
  author  = {Dagdelen, John and Dunn, Alexander and Lee, Sanghoon and others},
  title   = {{Structured information extraction from scientific text with large language models}},
  number  = {1},
  pages   = {1418},
  note    = {doi: \href{https://doi.org/10.1038/s41467-024-45563-x}{10.1038/s41467-024-45563-x}},
  volume  = {15},
  journal = {Nature Communications},
  year    = {2024}
}

@article{bubela_science-communication_2009,
  author  = {Bubela, Tania and Nisbet, Matthew and Borchelt, Rick and others},
  title   = {{Science communication reconsidered}},
  number  = {6},
  pages   = {514--518},
  note    = {doi: \href{https://doi.org/10.1038/nbt0609-514}{10.1038/nbt0609-514}},
  volume  = {27},
  journal = {Nature Biotechnology},
  year    = {2009}
}

@article{bruin_assessing-what_2013,
  author  = {Bruine de Bruin, Wändi and Bostrom, Ann},
  title   = {{Assessing what to address in science communication}},
  note    = {doi: \href{https://doi.org/10.1073/pnas.1212729110}{10.1073/pnas.1212729110}},
  pages   = {14062--14068},
  volume  = {110},
  journal = {Proceedings of the National Academy of Sciences},
  year    = {2013}
}

@article{Boy04,
  author  = {Boyack, Kevin},
  title   = {{Mapping knowledge domains: Characterizing PNAS}},
  pages   = {5192--5199},
  note    = {doi: \href{https://doi.org/10.1073/pnas.0307509100}{10.1073/pnas.0307509100}},
  volume  = {101},
  journal = {Proceedings of the National Academy of Sciences},
  year    = {2004}
}

@article{august_paper-plain_2023,
  author  = {August, Tal and Wang, Lucy and Bragg, Jonathan and others},
  title   = {{Paper Plain: Making Medical Research Papers Approachable to Healthcare Consumers with Natural Language Processing}},
  number  = {5},
  pages   = {1--38},
  note    = {doi: \href{https://doi.org/10.1145/3589955}{10.1145/3589955}},
  volume  = {30},
  journal = {ACM Transactions on Computer-Human Interaction},
  year    = {2023}
}

@article{airoldi_reconceptualizing-the_2010,
  author  = {Airoldi, Edoardo and Erosheva, Elena and Fienberg, Stephen and others},
  title   = {{Reconceptualizing the classification of PNAS articles}},
  number  = {49},
  pages   = {20899--20904},
  note    = {doi: \href{https://doi.org/10.1073/pnas.1013452107}{10.1073/pnas.1013452107}},
  volume  = {107},
  journal = {Proceedings of the National Academy of Sciences},
  year    = {2010}
}

@misc{Ope24,
  author = {{OpenAI}},
  title  = {{text-embedding-3-small}},
  note   = {{OpenAI Platform Website. Available at \url{https://platform.openai.com/docs/models/text-embedding-3-small/}. Accessed 9 February 2025.}},
  year   = {2024}
}

@misc{Nat25,
  author = {{National Academy of Sciences}},
  title  = {{PNAS home}},
  note   = {{Proceedings of the National Academy of Sciences Website. Available at \url{https://www.pnas.org/}. Accessed 3 January 2025.}},
  year   = {2025}
}

@misc{Ope241,
  author = {{OpenAI}},
  title  = {{GPT-4o mini}},
  note   = {{OpenAI Platform Website. Available at \url{https://platform.openai.com/docs/models/gpt-4o-mini}. Accessed 9 February 2025.}},
  year   = {2024}
}

@article{Nee22,
  author  = {Neelakantan, Arvind and Xu, Tao and Puri, Raul and others},
  title   = {{Text and Code Embeddings by Contrastive Pre-Training}},
  note    = {doi: \href{https://doi.org/10.48550/arXiv.2201.10005}{10.48550/arXiv.2201.10005}},
  year    = {2022},
  journal = {arXiv preprint arXiv:2201.10005}
}

@inproceedings{Bro20,
  author    = {Brown, Tom and Mann, Benjamin and Ryder, Nick and others},
  title     = {{Language Models are Few-Shot Learners}},
  pages     = {1877--1907},
  volume    = {33},
  booktitle = {Proceedings of the 34th International Conference on Neural Information Processing Systems, NIPS},
  address   = {Online},
  publisher = {Advances in Neural Information Processing Systems},
  year      = {2020}
}

@unpublished{Rad19,
  author = {Radford, Alec and Wu, Jeffrey and Child, Rewon and others},
  title  = {{Language Models are Unsupervised Multitask Learners}},
  url    = {https://cdn.openai.com/better-language-models/language_models_are_unsupervised_multitask_learners.pdf},
  year   = {2019},
  note   = {OpenAI preprint. Accesssed 14 November 2024}
}

@inproceedings{Fer23,
  author    = {Ferrando, Javier and Gállego, Gerald and Tsiamas, Ioannis and Costa-jussà, Marta},
  title     = {{Explaining How Transformers Use Context to Build Predictions}},
  pages     = {5486--5513},
  note      = {doi: \href{https://doi.org/10.18653/v1/2023.acl-long.301}{10.18653/v1/2023.acl-long.301}},
  volume    = {1},
  booktitle = {Proceedings of the 61st Annual Meeting of the Association for Computational Linguistics, ACL},
  address   = {Toronto, Canada},
  publisher = {Association for Computational Linguistics},
  year      = {2023}
}

@article{Mue23,
  author  = {Muennighoff, Niklas and Tazi, Nouamane and Magne, Loïc and Reimers, Nils},
  title   = {{MTEB: Massive Text Embedding Benchmark}},
  note    = {doi: \href{https://doi.org/10.48550/arXiv.2210.07316}{10.48550/arXiv.2210.07316}},
  year    = {2023},
  journal = {arXiv preprint arXiv:2210.07316}
}

@inproceedings{Qad19,
  author    = {Qader, Wisam and Ameen, Musa and Ahmed, Bilal},
  title     = {{An Overview of Bag of Words;Importance, Implementation, Applications, and Challenges}},
  pages     = {200--204},
  note      = {doi: \href{https://doi.org/10.1109/IEC47844.2019.8950616}{10.1109/IEC47844.2019.8950616}},
  booktitle = {Proceedings of the International Engineering Conference, IEC},
  address   = {Erbil, Iraq},
  publisher = {IEEE},
  year      = {2019}
}

@article{Sah23,
  author  = {Saha, Punyajoy and Garimella, Kiran and Kalyan, Narla and others},
  title   = {{On the rise of fear speech in online social media}},
  note    = {doi: \href{https://doi.org/10.1073/pnas.2212270120}{10.1073/pnas.2212270120}},
  number  = {11},
  pages   = {e2212270120},
  volume  = {120},
  journal = {Proceedings of the National Academy of Sciences},
  year    = {2023}
}

@article{Hah17,
  author  = {Hahsler, Michael and Karpienko, Radoslaw},
  title   = {{Visualizing association rules in hierarchical groups}},
  pages   = {317--335},
  note    = {doi: \href{https://doi.org/10.1007/s11573-016-0822-8}{10.1007/s11573-016-0822-8}},
  volume  = {87},
  journal = {Journal of Business Economics},
  year    = {2017}
}

@article{Ibr22,
  author  = {Ibrihich, Sara and Oussous, Ahmed and Ibrihich, Ouafaa and Esghir, Mustapha},
  title   = {{A Review on recent research in information retrieval}},
  note    = {doi: \href{https://doi.org/10.1016/j.procs.2022.03.106}{10.1016/j.procs.2022.03.106}},
  pages   = {777--782},
  volume  = {201},
  journal = {Procedia Computer Science},
  year    = {2022}
}

@inproceedings{Kot15,
  author    = {Kötter, Tobias and Günnemann, Stephan and Berthold, Michael and Faloutsos, Christos},
  title     = {{Automatic Taxonomy Extraction from Bipartite Graphs}},
  note      = {doi: \href{https://doi.org/10.1109/ICDM.2015.24}{10.1109/ICDM.2015.24}},
  pages     = {221--230},
  booktitle = {Proceedings of the IEEE International Conference on Data Mining},
  address   = {Atlantic City, NJ, USA},
  publisher = {IEEE},
  year      = {2015}
}

@article{van08,
  author  = {van der Maaten, Laurens and Hinton, Geoffrey},
  title   = {{Visualizing Data using t-SNE}},
  number  = {86},
  pages   = {2579--2605},
  url     = {http://jmlr.org/papers/v9/vandermaaten08a.html},
  volume  = {9},
  journal = {Journal of Machine Learning Research},
  year    = {2008}
}

@article{sme24,
  author  = {Smetana, Mason and Salles de Salles, Lucio and Sukharev, Igor and Khazanovich, Lev},
  title   = {{Highway Construction Safety Analysis Using Large Language Models}},
  number  = {4},
  pages   = {1352},
  note    = {doi: \href{https://doi.org/10.3390/app14041352}{https://doi.org/10.3390/app14041352}},
  volume  = {14},
  journal = {Applied Sciences},
  year    = {2024}
}

@article{egger_a-topic_2022,
  author  = {Egger, Roman and Yu, Joanne},
  title   = {{A Topic Modeling Comparison Between LDA, NMF, Top2Vec, and BERTopic to Demystify Twitter Posts}},
  journal = {Frontiers in Sociology},
  volume  = {7},
  year    = {2022},
  pages   = {886498},
  note    = {doi: \href{https://doi.org/10.3389/fsoc.2022.886498}{10.3389/fsoc.2022.886498}}
}

@article{ma_ai-powered_2025,
  author  = {Ma, Li and Chen, Ru and Ge, Weigong and others},
  title   = {{AI-powered topic modeling: comparing LDA and BERTopic in analyzing opioid-related cardiovascular risks in women}},
  volume  = {250},
  pages   = {10389},
  journal = {Experimental Biology and Medicine},
  year    = {2025},
  note    = {doi: \href{https://doi.org/10.3389/ebm.2025.10389}{10.3389/ebm.2025.10389}}
}

@misc{kaur_moving-beyond_2024,
  author  = {Kaur, Amandeep and Wallace, James R.},
  title   = {{Moving Beyond LDA: A Comparison of Unsupervised Topic Modelling Techniques for Qualitative Data Analysis of Online Communities}},
  note    = {doi: \href{https://doi.org/10.48550/arXiv.2412.14486}{10.48550/arXiv.2412.14486}},
  year    = {2024},
  journal = {arXiv preprint arXiv:2412.14486}
}

@article{benz_mapping-the_2025,
  author  = {Benz, Pierre and Pradier, Carolina and Kozlowski, Diego and Shokida, Natsumi S. and Larivière, Vincent},
  title   = {{Mapping the unseen in practice: comparing latent Dirichlet allocation and BERTopic for navigating topic spaces}},
  volume  = {130},
  pages   = {3839--3870},
  number  = {7},
  journal = {Scientometrics},
  year    = {2025},
  note    = {doi: \href{https://doi.org/10.1007/s11192-025-05339-6}{10.1007/s11192-025-05339-6}}
}

@inproceedings{hoyle_is-automated_2021,
  author    = {Hoyle, Alexander and Goel, Pranav and Hian-Cheong and others},
  title     = {{Is Automated Topic Model Evaluation Broken? The Incoherence of Coherence}},
  volume    = {34},
  url       = {https://proceedings.neurips.cc/paper_files/paper/2021/hash/0f83556a305d789b1d71815e8ea4f4b0-Abstract.html},
  pages     = {2018--2033},
  booktitle = {{Proceedings of the 35th Conference on Neural Information Processing Systems, NuerIPS}},
  publisher = {Curran Associates, Inc.},
  address   = {Online},
  year      = {2021}
}

@article{zhao_a-heuristic_2015,
  author  = {Zhao, Weizhong and Chen, James J and Perkins, Roger and others},
  title   = {A heuristic approach to determine an appropriate number of topics in topic modeling},
  volume  = {16},
  issn    = {1471-2105},
  pages   = {S8},
  issue   = {S13},
  journal = {{BMC Bioinformatics}},
  year    = {2015},
  note    = {doi: \href{https://doi.org/10.1186/1471-2105-16-S13-S8}{10.1186/1471-2105-16-S13-S8}}
}

@inproceedings{roder_exploring-the_2015,
  author    = {Röder, Michael and Both, Andreas and Hinneburg, Alexander},
  title     = {{Exploring the Space of Topic Coherence Measures}},
  isbn      = {978-1-4503-3317-7},
  series    = {{WSDM} '15},
  pages     = {399--408},
  booktitle = {{Proceedings of the Eighth ACM International Conference on Web Search and Data Mining}},
  publisher = {Association for Computing Machinery},
  year      = {2015},
  address   = {New York, {NY}, {USA}},
  url       = {https://doi.org/10.1145/2684822.2685324}
}

@inproceedings{doogan_topic-model_2021,
  author    = {Doogan, Caitlin and Buntine, Wray},
  address   = {Online},
  title     = {{Topic Model or Topic Twaddle? Re-evaluating Semantic Interpretability Measures}},
  url       = {https://doi.org/10.18653/v1/2021.naacl-main.300},
  pages     = {3824--3848},
  booktitle = {{Proceedings of the 2021 Conference of the North American Chapter of the Association for Computational Linguistics: Human Language Technologies}},
  publisher = {Association for Computational Linguistics},
  year      = {2021}
}

@article{theo_2025,
  author  = {Theocharopoulos, Panagiotis C. and Anagnostou, Panagiotis and Georgakopoulos, Spiros V. and Tasoulis, Sotiris K. and Plagianakos, Vassilis P.},
  title   = {{Large language models for efficient topic modeling}},
  volume  = {37},
  issn    = {1433-3058},
  pages   = {24421--24439},
  number  = {29},
  journal = {Neural Computing and Applications},
  year    = {2025},
  note    = {doi: \href{https://doi.org/10.1007/s00521-025-11593-9}{10.1007/s00521-025-11593-9}}
}

@inproceedings{pham_topicgpt_2024,
  author    = {Pham, Chau Minh and Hoyle, Alexander and Sun, Simeng and Resnik, Philip and Iyyer, Mohit},
  title     = {{TopicGPT: A Prompt-based Topic Modeling Framework}},
  booktitle = {{Proceedings of the 2024 Conference of the North American Chapter of the Association for Computational Linguistics: Human Language Technologies (Volume 1: Long Papers)}},
  pages     = {2956--2984},
  year      = {2024},
  url       = {https://aclanthology.org/2024.naacl-long.164/},
  address   = {Mexico City, Mexico},
  publisher = {Association for Computational Linguistics}
}

@misc{li_scitopic-enhancing_2025,
  author  = {Li, Pengjiang and Wang, Zaitian and Zhang, Xinhao and others},
  title   = {{SciTopic: Enhancing Topic Discovery in Scientific Literature through Advanced LLM}},
  year    = {2025},
  journal = {arXiv preprint arxiv:2508.20514},
  note    = {doi: \href{https://doi.org/10.48550/arXiv.2508.20514}{10.48550/arXiv.2508.20514}}
}

@article{hernandez_leveraging_2025,
  author  = {Hernandez-Camero, Ines and Garcia-Cabot, Antonio and Garcia-Lopez, Eva and Caro-Alvaro, Sergio and Moreno-Cediel, Antonio},
  title   = {{Leveraging large language models to enhance clustering-based topic modeling}},
  volume  = {67},
  issn    = {0219-1377, 0219-3116},
  pages   = {12661--12697},
  number  = {12},
  journal = {Knowledge and Information Systems},
  year    = {2025},
  note    = {doi: \href{https://doi.org/10.1007/s10115-025-02605-0}{10.1007/s10115-025-02605-0}}
}

@article{wu_survey_2024,
  author  = {Wu, Xiaobao and Nguyen, Thong and Luu, Anh Tuan},
  title   = {{A survey on neural topic models: methods, applications, and challenges}},
  volume  = {57},
  issn    = {1573-7462},
  pages   = {18},
  number  = {2},
  journal = {Artificial Intelligence Review},
  year    = {2024},
  note    = {doi: \href{https://doi.org/10.1007/s10462-023-10661-7}{10.1007/s10462-023-10661-7}}
}

@misc{mu_large_2024,
  author  = {Mu, Yida and Dong, Chun and Bontcheva, Kalina and Song, Xingyi},
  title   = {{Large Language Models Offer an Alternative to the Traditional Approach of Topic Modeling}},
  year    = {2024},
  journal = {arXiv preprint arxiv:2403.16248},
  note    = {doi: \href{https://doi.org/10.48550/arXiv.2403.16248}{10.48550/arXiv.2403.16248}}
}

@inproceedings{mimno_optimizing_2011,
  author    = {Mimno, David and Wallach, Hanna and Talley, Edmund and Leenders, Miriam and {McCallum}, Andrew},
  address   = {Edinburgh, Scotland, {UK}.},
  title     = {{Optimizing Semantic Coherence in Topic Models}},
  url       = {https://aclanthology.org/D11-1024/},
  pages     = {262--272},
  booktitle = {Proceedings of the 2011 Conference on Empirical Methods in Natural Language Processing},
  publisher = {Association for Computational Linguistics},
  year      = {2011}
}

@article{vaneck_10,
  author  = {Van Eck, N.J. and Waltman, L.},
  year    = {2010},
  title   = {{Software survey: VOSviewer, a computer program for bibliometric mapping}},
  journal = {Scientometrics},
  volume  = {84},
  pages   = {523--538},
  note    = {doi: \href{https://doi.org/10.1007/s11192-009-0146-3}{10.1007/s11192-009-0146-3}}
}

@article{donthu_21,
  author  = {Donthu, A. and Kumar, S. and Debmalya, M. and Pandey, N. and Lim, W.M.},
  year    = {2021},
  title   = {How to conduct a bibliometric analysis: {A}n overview and guidelines},
  journal = {Journal of Business Research},
  volume  = {133},
  pages   = {285--296},
  note    = {doi: \href{https://doi.org/10.1016/j.jbusres.2021.04.070}{10.1016/j.jbusres.2021.04.070}}
}

@article{weng_22,
  author  = {Weng, Min-Hsien and Wu, Shaoqun and Dyer, Mark},
  year    = {2022},
  title   = {Identification and Visualization of Key Topics in Scientific Publications with Transformer-Based Language Models and Document Clustering Methods},
  journal = {Applied Sciences},
  volume  = {12},
  number  = {21},
  pages   = {11220},
  note    = {doi: \href{https://doi.org/10.3390/app122111220}{10.3390/app122111220}}
}

@article{blei03,
  author  = {Blei, David and Ng, Andrew. and Jordan, Michael},
  year    = {2003},
  title   = {{Latent Dirichlet Allocation}},
  journal = {Journal of Machine Learning Research},
  volume  = {3},
  pages   = {993--1022},
  note    = {URL \href{https://dl.acm.org/doi/10.5555/944919.944937}{https://dl.acm.org/doi/10.5555/944919.944937}}
}

@article{lee_learning_1999,
  author  = {Lee, Daniel and Seung, H. Sebastian},
  year    = {1999},
  title   = {Learning the parts of objects by non-negative matrix factorization},
  journal = {Nature},
  volume  = {401},
  pages   = {788--791},
  note    = {doi: \href{https://doi.org/10.1038/44565}{10.1038/44565}}
}

@inproceedings{tran_topic-cropping_2013,
  author    = {Tran, Nam Khanh and Zerr, Sergej and Bischoff, Kerstin and Niederée, Claudia and Krestel, Ralf},
  title     = {{Topic Cropping: Leveraging Latent Topics for the Analysis of Small Corpora}},
  booktitle = {Research and Advanced Technology for Digital Libraries},
  year      = {2013},
  volume    = {8092},
  pages     = {297--308},
  publisher = {Springer},
  address   = {Berlin, Heidelberg},
  note      = {doi: \href{https://doi.org/10.1007/978-3-642-40501-3_30}{10.1007/978-3-642-40501-3\_30}}
}

\clearpage
\appendix

\section*{Appendix}
\setcounter{subsection}{0}

\subsection*{Abstract Classification Prompts}\label{sec:app_A}

For each $c_j^t \in C^t$, a subset of the most representative abstracts is selected to provide input for thematic generation. Representativeness is determined by calculating the cosine similarity between each abstract's embedding $\mathbb{E}(x_i \in c_j^t)$ and the corresponding cluster centroid, derived from K-means. This defines the subset of the top $k$ abstracts for cluster $c_j^t$, to be concatenated and used as the input prompt for the LLM $f(\cdot)$ to generate a theme label $T_j^t$ and description $D_j^t$:

\vspace{1em}

\begin{tcolorbox}[
colback=white, colframe=black, 
title=Summarize Topics Prompt,
coltitle=white,
colbacktitle=black,
fonttitle=\bfseries,
width=\textwidth,
boxrule=0.8pt,
arc=2mm,
left=3mm,
right=3mm]

The following is a list of paraphrased quotes from multiple papers:

\vspace{1ex}
\textbf{Quotes:} \{text\}

\vspace{1ex}
Take these and distill them into a comprehensive summary of the main ideas, methodology, and findings of the papers without leaving out any information. Also, generate 5 keywords (1 or 2 words each) and a single all-inclusive title (a few words) that may represent these papers. Provide your output as JSON as exemplified in the following format:

\vspace{1ex}
\hrule
\vspace{1ex}

\begin{verbatim}
Example Output:

{"summary": "distilled summary",
"title": "comprehensive title",
"keywords": ["keyword1", "keyword2", "keyword3", ...]}
\end{verbatim}
\end{tcolorbox}

\vspace{1em}

The following prompt takes keywords and titles from the first prompt for each cluster to generate brief descriptions $D_j^t$. These two prompts are intentionally separated to consolidate information and reduce the length of injected context during LLM inference. If the model is provided too much context during inference, it may struggle to accurately summarize the information and get “lost in the middle” \cite{Liu24}. In this prompt, only one output example is shown for conceptual simplification, but this can be expanded to multiple based on user preferences.

\vspace{1em}

\begin{tcolorbox}[
colback=white, colframe=black, 
title=Generate Classes Prompt,
coltitle=white,
colbacktitle=black,
fonttitle=\bfseries,
width=\textwidth,
boxrule=0.8pt,
arc=2mm,
left=3mm,
right=3mm]

Create a JSON output similar to the example based on the following titles and keywords:

\vspace{1ex}
\textbf{Titles and Keywords:} \{text\}

\vspace{1ex}
Each class should have a brief description and a representative title that is only 2-3 words long.

\vspace{1ex}
\hrule
\vspace{1ex}

\begin{verbatim}
Example Output:

{"classes": [{
    "class": "0",
    "title": "Investment and Funding",
    "desc": "Allocation of financial resources to ensure 
    equitable transportation infrastructure and services, 
    focusing on underserved and communities."
}, AND SO ON...]}
\end{verbatim}
\end{tcolorbox}

\vspace{1em}

\clearpage
\subsection*{LLM-Based Reclassification}

This formulation represents a conditional classification task, where the LLM selects the most appropriate category $j$ for each abstract based on the similarity between the content of $x_i$ and the semantic meaning conveyed by each ($T_j^t$, $D_j^t$) pair. By leveraging its self-attention mechanisms and pretrained knowledge \cite{Vas17}, the LLM can assess complex relationships between the abstract text and the thematic descriptions, thereby refining the classification beyond the initial unsupervised clustering. This step ensures that each abstract is explicitly aligns with a thematic category, as interpreted and articulated by the model. Similar to the prompts utilized to generate thematic categories ($T_j^t$, $D_j^t$), the prompt used to reclassify abstracts and later used for full-text classification is as follows:

\vspace{1em}

\begin{tcolorbox}[
colback=white, colframe=black, 
title=Classify with JSON,
coltitle=white,
colbacktitle=black,
fonttitle=\bfseries,
width=\textwidth,
boxrule=0.8pt,
arc=2mm,
left=3mm,
right=3mm]

You are an expert in general multi-class text classification tasks. Classify the provided text into an appropriate category using the following descriptions of possible categories.

\vspace{1ex}
\textbf{Categories:} \{classes\}

\textbf{Provided Text:} \{text\}

\vspace{1ex}
Report your output as JSON as exemplified below. If no categories seem appropriate for the text select the "Other" classification. Report the associated id in the "\{id\}" key and the predicted category in the "\{class\}" key FOR EACH UNIQUE ID in the provided text. 

\vspace{1ex}
\hrule
\vspace{1ex}

\begin{verbatim}
Example Categories:

{"classes": [{
    "class": "0",
    "title": "Technology",
    "desc": "Articles related to the latest technological 
    advancements, gadgets, software, and hardware."
}, AND SO ON...]}
___
Example Output:

{"chunks": [
    {"id": "001", "{class}": "0"},
    {"id": "384", "{class}": "2"},
    AND SO ON...]
}
\end{verbatim}
\end{tcolorbox}

\vspace{1em}

\clearpage
\begin{figure}
\centering
\includegraphics[width=\textwidth]{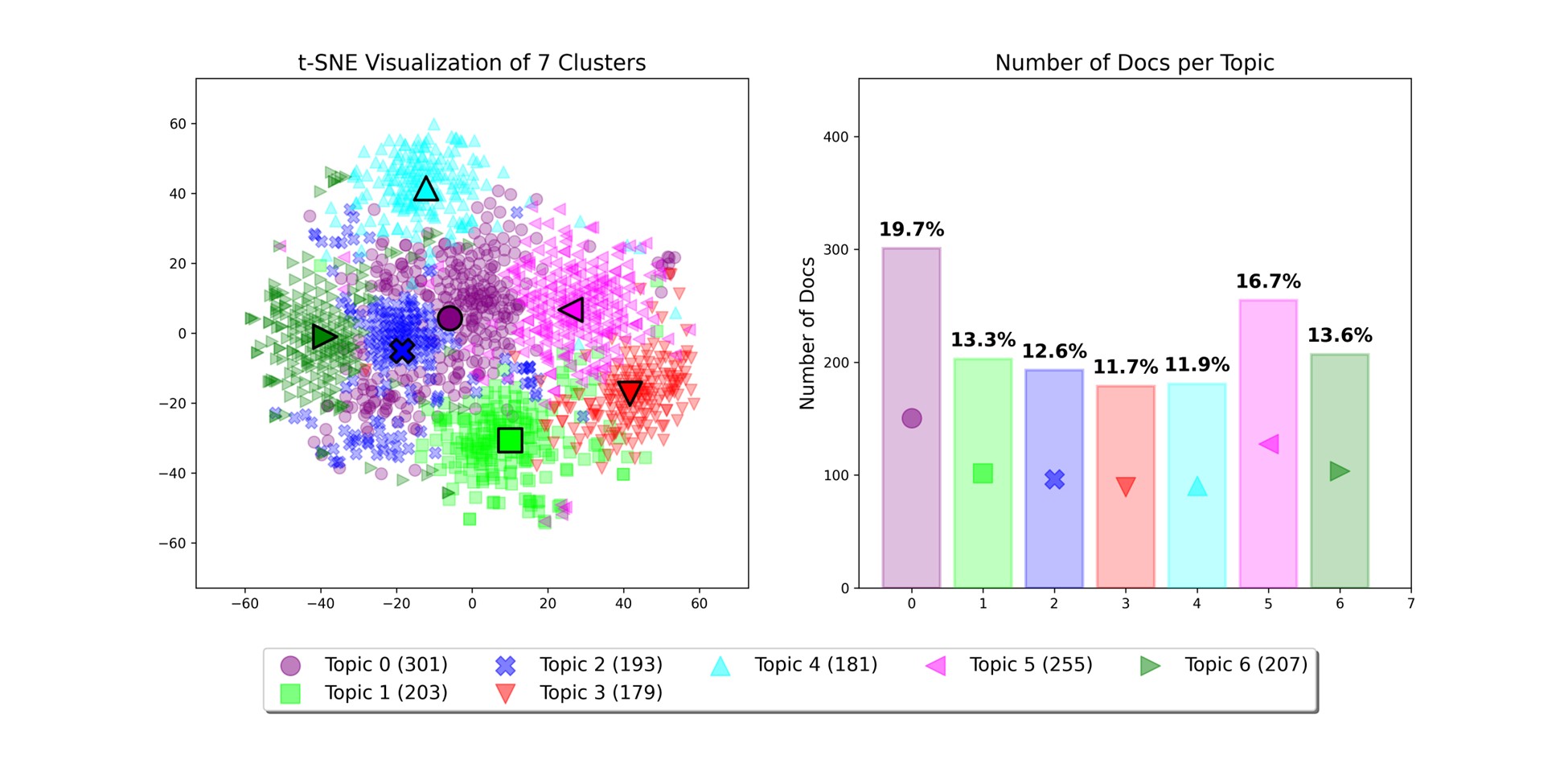}
\caption{Example abstract K-means clustering for iteration $t=1$ with 1,519 abstracts and $k=7$ clusters. The representation on the left-hand side of the plot is a 2D t-distributed stochastic neighbor embedding (t-SNE) projection of embedding vectors of length 1,536 \cite{van08} The right-hand side shows the distribution of abstracts between clusters, with the highest belonging to Topic 0 at 19.7\%.}
\label{fig:kmeans}
\end{figure}

\clearpage
\begin{figure}
\centering
\includegraphics[width=\textwidth]{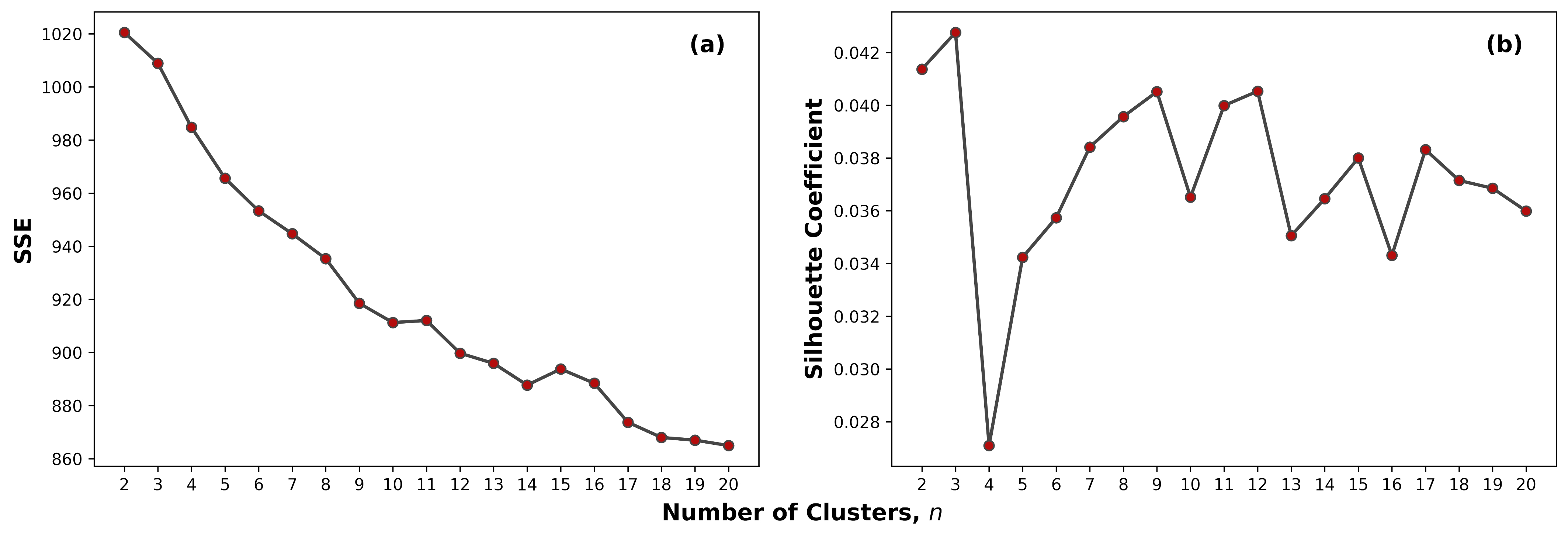}
\caption{Varying K-means clustering over 1,519 PNAS abstracts from $k=2$ to $k=20$ clusters. Plot (a) shows the number of clusters $k$ versus the average cluster-wise Sum of Squared Error (SSE). Plot (b) similarly shows $k$ versus the Silhouette coefficient. Neither plot indicates an optimal number of clusters from K-means alone.}
\label{fig:nclusters}
\end{figure}

\clearpage
\begin{figure}
\centering
\includegraphics[width=\textwidth]{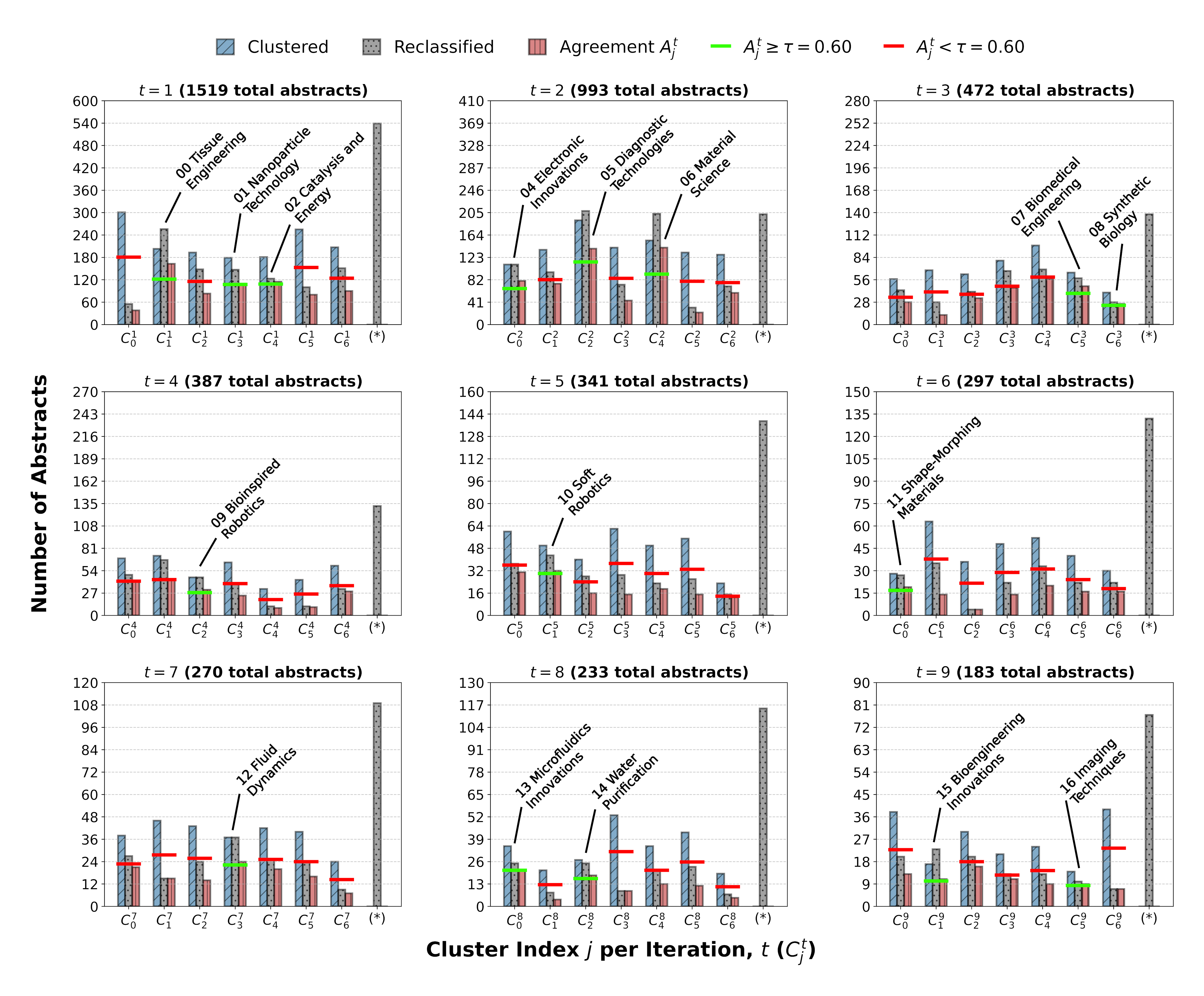}
\caption{All nine iterations of primary abstract classification. Each subplot shows interim topics: the number of clustered and reclassified abstracts. The $\tau = 0.60$ agreement threshold is also plotted to indicate which interim topics were considered stable and preserved from subsequent iterations. These stable topics were labeled with their associated LLM-generated title ($T_J$) in the order in which they appeared.}
\label{fig:interim_topics}
\end{figure}

\clearpage
\begin{figure}
\centering
\includegraphics[width=\textwidth]{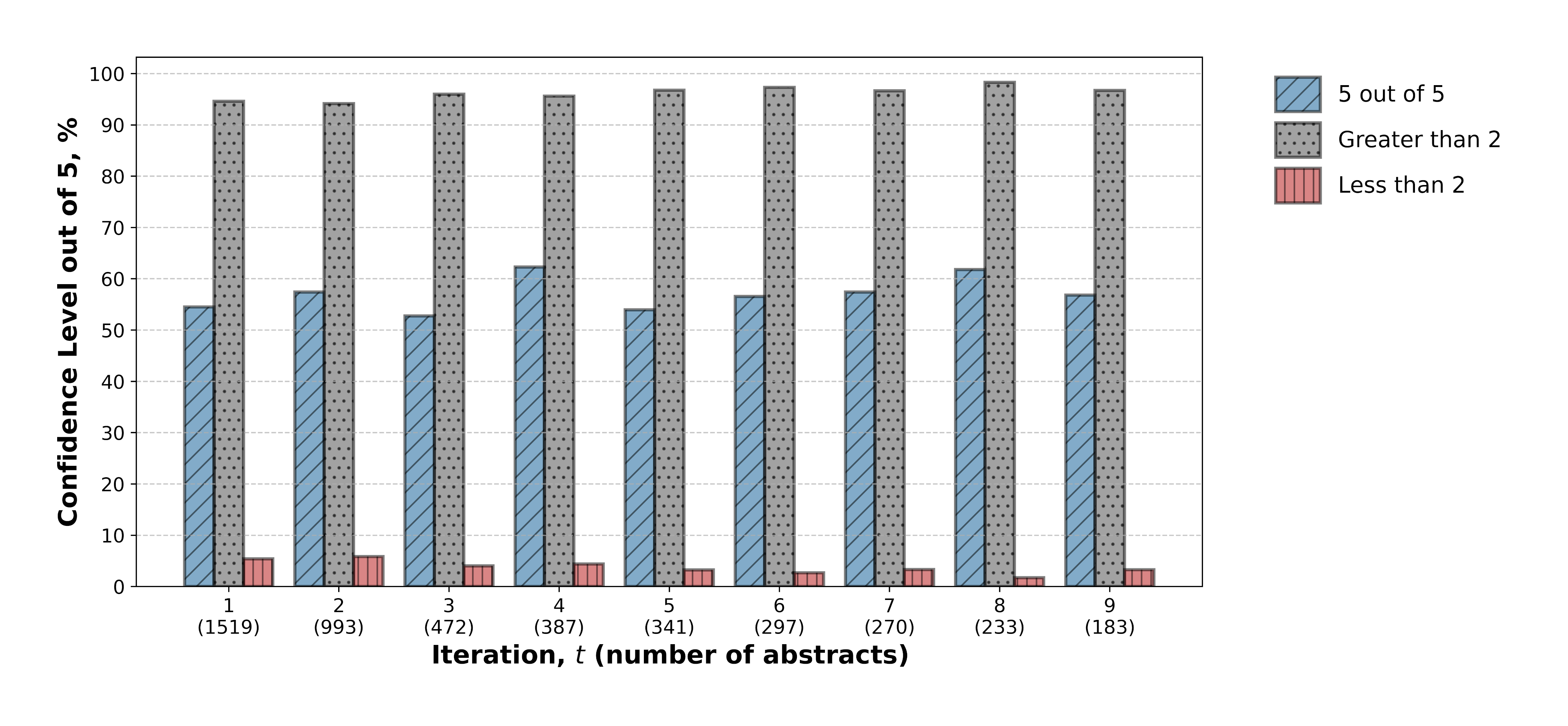}
\caption{All nine iterations of primary abstract classification, and cases in which the abstract satisfied the “3-out-of-5” criteria.}
\label{fig:3of5_criteria}
\end{figure}

\clearpage
\begin{table}
\small
\caption{Complete breakdown of published articles in PNAS subcategories under biological, physical, and social sciences (in descending order) from 2005 to 2024. Adapted from the PNAS website \cite{Nat25}. The Engineering category is emphasized to illustrate the journal's contribution relative to other subcategories.}
\begin{tabularx}{\textwidth}{@{\extracolsep{\fill}} l|rrrr|rr }
\toprule
& 2005 & 2010 & 2015 & 2020 & 2005 & \\
Category & -2009 & -2014 & -2019 & -2024 & -2024 & \% Pub.\\
\midrule
1. Neuroscience & 2,008 & 2,109 & 1,699 & 1,560 & 7,376 & 9.74\% \\
2. Biochemistry & 2,002 & 1,902 & 1,424 & 1,103 & 6,431 & 8.49\% \\
3. Biophysics and Computational Biology & 1,611 & 1,571 & 1,248 & 1,155 & 5,585 & 7.37\% \\
4. Medical Sciences & 1,738 & 1,501 & 1,036 & 718 & 4,993 & 6.59\% \\
5. Microbiology & 1,081 & 1,215 & 1,046 & 1,156 & 4,498 & 5.94\% \\
6. Cell Biology & 1,268 & 1,296 & 906 & 902 & 4,372 & 5.77\% \\
7. Immunology and Inflammation & 1,108 & 1,223 & 902 & 912 & 4,145 & 5.47\% \\
8. Evolution & 843 & 813 & 748 & 690 & 3,094 & 4.09\% \\
9. Genetics & 878 & 769 & 587 & 505 & 2,739 & 3.62\% \\
10. Plant Biology & 564 & 643 & 609 & 546 & 2,362 & 3.12\% \\
11. Ecology & 444 & 480 & 574 & 612 & 2,110 & 2.79\% \\
12. Developmental Biology & 605 & 552 & 365 & 348 & 1,870 & 2.47\% \\
13. Physiology & 605 & 552 & 365 & 348 & 1,870 & 2.47\% \\
14. Applied Biological Sciences & 306 & 372 & 303 & 321 & 1,302 & 1.72\% \\
15. Psychological and Cognitive Sciences & 159 & 234 & 224 & 213 & 830 & 1.10\% \\
16. Environmental Sciences & 134 & 255 & 202 & 189 & 780 & 1.03\% \\
17. Pharmacology & 207 & 188 & 143 & 123 & 661 & 0.87\% \\
18. Anthropology & 110 & 164 & 134 & 120 & 528 & 0.70\% \\
19. Systems Biology & 10 & 192 & 152 & 114 & 468 & 0.62\% \\
20. Agricultural Sciences & 88 & 103 & 106 & 155 & 452 & 0.60\% \\
21. Sustainability Science & 66 & 96 & 103 & 114 & 379 & 0.50\% \\
22. Population Biology & 57 & 54 & 64 & 97 & 272 & 0.36\% \\
\textbf{All Biological Sciences Subcategories} & \textbf{15,699} & \textbf{16,104} & \textbf{12,879} & \textbf{11,949} & \textbf{56,631} & \textbf{74.8\%} \\
\midrule
1. Chemistry & 868 & 984 & 866 & 944 & 3,662 & 4.83\% \\
2. Earth, Atmospheric, and Planetary Sciences & 224 & 496 & 633 & 765 & 2,118 & 2.80\% \\
3. Applied Physical Sciences & 294 & 541 & 588 & 664 & 2,087 & 2.76\% \\
4. Physics & 246 & 462 & 542 & 556 & 1,806 & 2.38\% \\
\rowcolor{gray!20} 
\textbf{5. Engineering} & \textbf{129} & \textbf{298} & \textbf{432} & \textbf{647} & \textbf{1,506} & \textbf{1.99\%} \\
6. Biophysics and Computational Biology & -- & -- & 387 & 754 & 1,141 & 1.51\% \\
7. Environmental Sciences & 117 & 238 & 218 & 222 & 795 & 1.05\% \\
8. Applied Mathematics & 154 & 157 & 153 & 179 & 643 & 0.85\% \\
9. Statistics & 53 & 66 & 89 & 89 & 297 & 0.39\% \\
10. Computer Sciences & 42 & 56 & 67 & 123 & 288 & 0.38\% \\
11. Mathematics & 50 & 82 & 37 & 34 & 203 & 0.27\% \\
12. Astronomy & 13 & 38 & 21 & 30 & 102 & 0.13\% \\
13. Sustainability Science & -- & -- & -- & 65 & 65 & 0.09\% \\
\textbf{All Physical Sciences Subcategories} & \textbf{2,190} & \textbf{3,418} & \textbf{4,033} & \textbf{5,072} & \textbf{14,713} & \textbf{19.4\%} \\
\midrule
1. Psychological and Cognitive Sciences & 153 & 377 & 568 & 559 & 1,657 & 2.19\% \\
2. Sustainability Science & 62 & 181 & 227 & 187 & 657 & 0.87\% \\
3. Social Sciences & 65 & 132 & 207 & 193 & 597 & 0.79\% \\
4. Anthropology & 101 & 139 & 179 & 143 & 562 & 0.74\% \\
5. Economic Sciences & 60 & 126 & 127 & 196 & 509 & 0.67\% \\
6. Environmental Sciences & 18 & 45 & 63 & 102 & 228 & 0.30\% \\
7. Political Sciences & 101 & 139 & 179 & 143 & 562 & 0.74\% \\
8. Demography & -- & -- & -- & 30 & 30 & 0.04\% \\
\textbf{All Social Sciences Subcategories} & \textbf{466} & \textbf{1,015} & \textbf{1,409} & \textbf{1,506} & \textbf{4,396} & \textbf{5.80\%} \\
\bottomrule
\end{tabularx}
\label{tab:pnas_pubcount}
\end{table}

\clearpage
\begin{table}
\small
\caption{Sixteen LLM-derived topics and distributions of initial clustering (“No. Cluster”) and reclassification (“No. Class”). $A_j^t$ represents the corresponding agreement score for iteration (“Iter.”) $t$, all greater than or equal to $\tau = 0.60$. The asterisk (*) in the final row distinguishes between LLM-generated descriptions and the default fallback description for the $k+1$ “Other” category.}
\begin{tabularx}{\textwidth}{@{\extracolsep{\fill}} lrrrr}
\toprule
Theme Label ($T_J$) & Iter. ($t$) & No. Cluster & No. Class & $A_j^t$ \\
\midrule
01 -- Tissue Engineering (and stem cell behavior) & 1 & 203 & 256 & 80.3\% \\
02 -- Nanoparticle Technology (for targeted cancer therapy) & 1 & 179 & 147 & 61.5\% \\
03 -- Catalysis and Energy Storage & 1 & 181 & 123 & 63.5\% \\
04 -- Electronic Innovations & 2 & 110 & 110 & 72.7\% \\
05 -- Diagnostic Technologies & 2 & 191 & 208 & 72.8\% \\
06 -- Material Science & 2 & 154 & 203 & 91.6\% \\
07 -- Biomedical Engineering & 3 & 65 & 58 & 73.8\% \\
08 -- Synthetic Biology & 3 & 40 & 28 & 65.0\% \\
09 -- Bioinspired Robotics & 4 & 46 & 46 & 67.4\% \\
10 -- Soft Robotics & 5 & 50 & 43 & 64.0\% \\
11 -- Shape-Morphing Materials & 6 & 28 & 27 & 67.9\% \\
12 -- Fluid Dynamics & 7 & 37 & 37 & 64.9\% \\
13 -- Microfluidics Innovation & 8 & 35 & 25 & 60.0\% \\
14 -- Water Purification & 8 & 27 & 25 & 66.7\% \\
15 -- Bioengineering Innovations & 9 & 17 & 23 & 64.7\% \\
16 -- Imaging Techniques & 9 & 14 & 10 & 64.3\% \\
17 -- Other* & -- & -- & 150 & -- \\
\midrule
\textbf{Total} & & & \textbf{1,519} & \\
\bottomrule
\end{tabularx}
\label{tab:cluster_agreement}
\end{table}

\vspace{0.5em}

\footnotesize
\textbf{Lay LLM Descriptions ($D_J$):}
\begin{enumerate}
\item Study of tissue engineering and mechanotransduction, emphasizing the role of extracellular matrix and stem cell behavior in biomaterials.
\item Advancements in cancer treatment utilizing nanoparticle technology for targeted drug delivery and personalized medicine.
\item Research on catalysis and energy storage technologies, focusing on CO$_2$ conversion and sustainable lithium battery solutions.
\item Exploration of cutting-edge electronic and photonic technologies, focusing on 2D materials and flexible electronics.
\item Innovations in diagnostic technologies aimed at enhancing healthcare outcomes through real-time analysis and personalized medicine.
\item Investigation of mechanical properties at micro and nanoscale levels, focusing on micromechanics and additive manufacturing.
\item Development of flexible and biointegrated technologies in biomedical engineering, focusing on wearable health monitoring solutions.
\item Innovations in synthetic biology, emphasizing tools and methodologies for gene regulation and metabolic processes.
\item Insights into bioinspired robotics, highlighting locomotion and collective behavior derived from natural systems.
\item Innovations in soft robotics, emphasizing advancements in material design and actuation mechanisms for biomedical applications.
\item Exploration of innovative materials and technologies that can change shape and function, enhancing applications in soft robotics and programmable matter.
\item Insights into fluid dynamics, including turbulent flow and viscoelastic fluids, with applications in active systems.
\item Exploration of cutting-edge advancements in microfluidics technology, focusing on fluid manipulation and particle analysis for various applications.
\item Development of advanced technologies for water purification, emphasizing membrane technology and sustainable wastewater treatment solutions.
\item Advancements in bioengineering and human-machine interfaces aimed at enhancing medical technology and sustainable agriculture.
\item Transformative advances in biological imaging methods, including 3D and Raman imaging for nanoscale resolution.
\item None of the other categories fit this text.
\end{enumerate}

\clearpage
\begin{table}
\small
\caption{Summary of secondary LLM (full-text) classification. For each topic, and therefore each abstract, due to multi-label tolerance, a significant number of segments were classified under more than one topic in $S^*$. The “Same” column indicates how many segments were classified under the same parent abstract topic ($u_{ik}=y_i$). The “To Other” column represents the number of segments that were classified to other topics aside from its parent abstract topic ($u_{ik} \subseteq S^*$, where $u_{ik} \neq y_i$). “No. Gained” shows the influx of classifications from other topics outside the parent abstract topic. Finally, the “\% Corpus” shows the overall percent of texts that were classified under a specific topic $T_J$ as the summation of the number of self-classifications ($u_{ik}=y_i$) and the number gained divided by the total number of text segment classifications (46,639).}
\begin{tabularx}{\textwidth}{@{\extracolsep{\fill}} l|rrr|rrrr|r}
\toprule
& No. & No. & No. & No. & Same & To Other & No.& \% \\
Topic ($T_J$) & Abs. & Seg. & Class & Exclusive & ($u_{ik}=y_i$) & ($u_{ik} \neq y_i$) & Gained & Corpus \\
\midrule
01 -- Tissue Engineering\ldots & 256 & 3,368 & 11,154 & 13 & 1,907 & 9,247 & 962 & 5.78\% \\
02 -- Nanoparticle Technology\ldots & 147 & 2,081 & 6,465 & 20 & 1,332 & 5,133 & 864 & 4.42\% \\
03 -- Catalysis and Energy & 123 & 1,325 & 2,918 & 110 & 931 & 1,987 & 330 & 2.54\% \\
04 -- Electronic Innovations & 110 & 1,172 & 2,494 & 289 & 841 & 1,653 & 2,665 & 7.06\% \\
05 -- Diagnostic Technologies & 208 & 2,798 & 7,452 & 292 & 2,131 & 5,321 & 4,344 & 13.0\% \\
06 -- Material Science & 203 & 2,136 & 4,495 & 664 & 1,817 & 2,678 & 3,561 & 10.8\% \\
07 -- Biomedical Engineering & 58 & 753 & 2,369 & 9 & 559 & 1,810 & 5,011 & 11.2\% \\
08 -- Synthetic Biology & 28 & 309 & 700 & 47 & 241 & 459 & 2,212 & 4.94\% \\
09 -- Bioinspired Robotics & 46 & 574 & 1,493 & 78 & 353 & 1,140 & 771 & 2.26\% \\
10 -- Soft Robotics & 43 & 551 & 2,183 & 1 & 368 & 1,815 & 1,886 & 4.54\% \\
11 -- Shape-Morphing Materials & 27 & 284 & 920 & 13 & 201 & 719 & 2,544 & 5.53\% \\
12 -- Fluid Dynamics & 37 & 336 & 732 & 79 & 291 & 441 & 2,415 & 5.45\% \\
13 -- Microfluidics Innovations & 25 & 295 & 990 & 5 & 186 & 804 & 1,978 & 4.36\% \\
14 -- Water Purification & 25 & 339 & 798 & 21 & 191 & 607 & 271 & 0.93\% \\
15 -- Bioengineering Innovations & 23 & 330 & 717 & 12 & 151 & 566 & 5,806 & 12.0\% \\
16 -- Imaging Techniques & 10 & 148 & 264 & 27 & 105 & 159 & 1,457 & 3.15\% \\
17 -- Other & 150 & 1,837 & 3,495 & 438 & 438 & 3,057 & 519 & 1.93\% \\
\midrule
\textbf{Sum} & \textbf{1,519} & \textbf{18,636} & \textbf{49,639} & \textbf{2,118} & \textbf{12,043} & \textbf{37,596} & \textbf{37,596} & \textbf{100\%} \\
\bottomrule
\end{tabularx}
\label{tab:fulltext_dist}
\end{table}

\clearpage
\begin{table}
\small
\caption{The adjacency matrix $M$ results from a $G=(U, V, E)$ bipartite graph representation for 16 stable topics ($S^*$). The rows represent the set of $U$ nodes from primary abstract classification, and the columns represent the set of $V$ nodes due to secondary full-text classification. The value in each cell represents the cumulative weights of edges $E$ between $U$ and $V$. The black shaded cells illustrate the most dominant connection between abstract and full text for all topics.}
\setlength{\tabcolsep}{2pt}
\begin{tabularx}{\textwidth}{@{\extracolsep{\fill}} c *{17}{r}}
\toprule
Topic ($T_J$) & 01 & 02 & 03 & 04 & 05 & 06 & 07 & 08 & 09 & 10 & 11 & 12 & 13 & 14 & 15 & 16 & 17 \\
\midrule
01 & \cellcolor{black}{\textcolor{white}{1,907}} & 168 & 1 & 143 & 1,820 & 1,061 & 1,870 & 499 & 69 & 552 & 366 & 250 & 340 & 11 & 1,776 & 262 & 59 \\
02 & 104 & 1,332 & 16 & 170 & \cellcolor{black}{\textcolor{white}{1,501}} & 219 & 771 & 673 & 23 & 107 & 115 & 80 & 156 & 9 & 928 & 220 & 41 \\
03 & 11 & 45 & \cellcolor{black}{\textcolor{white}{931}} & 497 & 34 & 528 & 70 & 78 & 19 & 15 & 72 & 150 & 49 & 126 & 227 & 29 & 37 \\
04 & 36 & 75 & 64 & \cellcolor{black}{\textcolor{white}{841}} & 34 & 195 & 179 & 17 & 63 & 98 & 277 & 100 & 107 & 3 & 209 & 155 & 41 \\
05 & 267 & 372 & 10 & 214 & \cellcolor{black}{\textcolor{white}{2,131}} & 152 & 1,173 & 597 & 21 & 122 & 54 & 245 & 560 & 33 & 895 & 533 & 73 \\
06 & 172 & 42 & 84 & 141 & 11 & \cellcolor{black}{\textcolor{white}{1,817}} & 183 & 66 & 81 & 241 & 575 & 521 & 122 & 11 & 274 & 81 & 73 \\
07 & 121 & 60 & 19 & 274 & 330 & 73 & \cellcolor{black}{\textcolor{white}{559}} & 30 & 40 & 186 & 124 & 13 & 44 & 5 & 441 & 19 & 31 \\
08 & 19 & 16 & 6 & 58 & 76 & 51 & 8 & \cellcolor{black}{\textcolor{white}{241}} & 14 & 22 & 26 & 16 & 7 & 2 & 123 & 2 & 13 \\
09 & 21 & 2 & 11 & 141 & 6 & 145 & 75 & 12 & \cellcolor{black}{\textcolor{white}{353}} & 184 & 171 & 150 & 31 & 9 & 132 & 9 & 41 \\
10 & 44 & 8 & 3 & 274 & 19 & 134 & 222 & 4 & 210 & 368 & \cellcolor{black}{\textcolor{white}{395}} & 143 & 77 & 1 & 266 & 1 & 14 \\
11 & 17 & 0 & 8 & 109 & 0 & 120 & 60 & 3 & 43 & 125 & \cellcolor{black}{\textcolor{white}{201}} & 86 & 43 & 11 & 86 & 3 & 5 \\
12 & 17 & 2 & 8 & 37 & 5 & 161 & 12 & 2 & 12 & 25 & 35 & \cellcolor{black}{\textcolor{white}{291}} & 83 & 3 & 15 & 13 & 11 \\
13 & 18 & 7 & 8 & 91 & 37 & 103 & 94 & 12 & 15 & 47 & 76 & 180 & \cellcolor{black}{\textcolor{white}{186}} & 12 & 83 & 11 & 10 \\
14 & 11 & 32 & 40 & 37 & 64 & 139 & 26 & 22 & 7 & 9 & 26 & 59 & 10 & \cellcolor{black}{\textcolor{white}{191}} & 95 & 17 & 13 \\
15 & 26 & 0 & 5 & 64 & 96 & 48 & 69 & 65 & 32 & 29 & 28 & 26 & 24 & 5 & \cellcolor{black}{\textcolor{white}{151}} & 9 & 40 \\
16 & 4 & 1 & 0 & 47 & 32 & 12 & 19 & 11 & 0 & 2 & 3 & 3 & 7 & 0 & 1 & \cellcolor{black}{\textcolor{white}{105}} & 17 \\
17 & 74 & 34 & 47 & 368 & 279 & 420 & 180 & 121 & 122 & 122 & 201 & 393 & 318 & 30 & 255 & 93 & \cellcolor{black}{\textcolor{white}{438}} \\
\midrule
\textbf{Sum} $\Sigma$ & 2,869 & 2,196 & 1,261 & 3,506 & 6,475 & 5,378 & 5,570 & 2,453 & 1,124 & 2,254 & 2,745 & 2,706 & 2,164 & 456 & 5,957 & 1,562 & 957 \\
\textbf{Percent} & 5.78 & 4.42 & 2.54 & 7.06 & 13.0 & 10.8 & 11.2 & 4.94 & 2.26 & 4.54 & 5.53 & 5.45 & 4.36 & 0.93 & 12.0 & 3.15 & 1.93 \\
\bottomrule
\end{tabularx}
\label{tab:adjaceny_mat}
\end{table}

\vspace{0.5em}

\footnotesize
\textbf{Theme Labels ($T_J$):}
\begin{itemize}
\item 01 -- Tissue Engineering (and stem cell behavior)
\item 02 -- Nanoparticle Technology (for targeted cancer therapy)
\item 03 -- Catalysis and Energy
\item 04 -- Electronic Innovations
\item 05 -- Diagnostic Technologies
\item 06 -- Material Science
\item 07 -- Biomedical Engineering
\item 08 -- Synthetic Biology
\item 09 -- Bioinspired Robotics
\item 10 -- Soft Robotics
\item 11 -- Shape-Morphing Materials
\item 12 -- Fluid Dynamics
\item 13 -- Microfluidics Innovations
\item 14 -- Water Purification
\item 15 -- Bioengineering  Innovations
\item 16 -- Imaging Techniques
\item 17 -- Other
\end{itemize}

\end{document}